\documentclass[journal]{IEEEtran}

\usepackage{cite}
\usepackage{graphicx}
\usepackage{amsmath}
\usepackage{amsfonts}
\usepackage{subfigure}
\usepackage{algorithmic}
\usepackage{algorithm}
\usepackage[pagewise]{lineno}

\newtheorem{remark}{Remark}

\begin{document}

\title{Joint Relay and Jammer Selection Improves the Physical Layer Security in the Face of CSI Feedback Delays}

\author{Lei Wang,~\IEEEmembership{Student Member,~IEEE},
        Yueming Cai,~\IEEEmembership{Senior Member,~IEEE},
        Yulong Zou,~\IEEEmembership{Senior Member,~IEEE},
        Weiwei Yang,~\IEEEmembership{Member,~IEEE},
        and~Lajos Hanzo,~\IEEEmembership{Fellow,~IEEE}
\thanks{Copyright (c) 2015 IEEE. Personal use of this material is permitted. However, permission to use this material for any other purposes must be obtained from the IEEE by sending a request to pubs-permissions@ieee.org.}
\thanks{Manuscript received October 2, 2014; revised February 25, 2015 and July 27, 2015; accepted September 8, 2015. This work was partially supported by the National Natural Science Foundation of China (Nos. 61371122, 61471393 and 61501512) and the Natural Science Foundation of Jiangsu Province (Nos. BK20150718 and BK20150040). The review of this paper was coordinated by Prof. M. Cenk Gursoy.}
\thanks{L. Wang, Y. Cai, and W. Yang are with the College of Communications Engineering, PLA University of Science and Technology, Nanjing 210007, China (Email: csu-wl@163.com, caiym@vip.sina.com, yww\_1010@aliyun.com).}
\thanks{Y. Zou is with the School of Telecommunications and Information Engineering, Nanjing University of Posts and Telecommunications, Nanjing 210003, China (E-mail: yulong.zou@njupt.edu.cn.).}
\thanks{L. Hanzo is with the Department of Electronics and Computer Science, University of Southampton, Southampton, United Kingdom (E-mail: lh@ecs.soton.ac.uk).}}

\markboth{IEEE Transactions on Vehicular Technology (Accepted for Publication)}
{L. Wang \MakeLowercase{\textit{et al.}}: Joint Relay and Jammer Selection Improves the Physical Layer Security in the Face of CSI Feedback Delays}

\maketitle

\begin{abstract}
We enhance the physical-layer security (PLS) of amplify-and-forward relaying networks with the aid of joint relay and jammer selection (JRJS), despite the deliterious effect of channel state information (CSI) feedback delays. Furthermore, we conceive a new outage-based characterization approach for the JRJS scheme. The traditional best relay selection (TBRS) is also considered as a benchmark. We first derive closed-form expressions of both the connection outage probability (COP) and of the secrecy outage probability (SOP) for both the TBRS and JRJS schemes. Then, a reliable-and-secure connection probability (RSCP) is defined and analyzed for characterizing the effect of the correlation between the COP and SOP introduced by the corporate source-relay link. The reliability-security ratio (RSR) is introduced for characterizing the relationship between the reliability and security through the asymptotic analysis. Moreover, the concept of effective secrecy throughput is defined as the product of the secrecy rate and of the RSCP for the sake of characterizing the overall efficiency of the system, as determined by the transmit SNR, secrecy codeword rate and the power sharing ratio between the relay and jammer. The impact of the direct source-eavesdropper link and additional performance comparisons with respect to other related selection schemes are further included. Our numerical results show that the JRJS scheme outperforms the TBRS method both in terms of the RSCP as well as in terms of its effective secrecy throughput, but it is more sensitive to the feedback delays. Increasing the transmit SNR will not always improve the overall throughput. Moreover, the RSR results demonstrate that upon reducing the CSI feedback delays, the reliability improves more substantially than the security degrades, implying an overall improvement in terms of the security-reliability tradeoff. Additionally, the secrecy throughput loss due to the second hop feedback delay is more pronounced than that of the first hop.
\end{abstract}


\section{Introduction}\label{sec1}

\IEEEPARstart{W}{ireless} communications systems are particularly vulnerable to security attacks because of the inherent openness of the transmission medium. Traditionally, the information privacy of wireless networks has been focused on the higher layers of the protocol stack employing cryptographically secure schemes. However, these methods typically assume a limited computing power for the eavesdroppers and exhibit inherent vulnerabilities in terms of the inevitable secret key distribution as well as management [1]. In recent years, physical-layer security (PLS) has emerged as a promising technique of improving the confidentiality wireless communications, which exploits the time varying properties of fading channels, instead of relying on conventional cryptosystems. The pivotal idea of PLS solutions is to exploit the dynamically fluctuating random nature of radio channels for maximizing the uncertainty concerning the source messages at the eavesdropper [2], [3].

To achieve this target, several PLS-enhancement approaches have been proposed in the literature, including secrecy-enhancing channel coding [4], secure on-off transmission designs [5], secrecy-improving beamforming/precoding and artificial noise (AN) aided techniques relying on multiple-antennas [6], as well as secure relay-assisted transmission techniques [7]. Specifically, apart from improving the reliability and coverage of wireless transmissions, user cooperation also has a great potential in terms of enhancing the wireless security against eavesdropping attacks. There has been a growing interest in improving the security of cooperative networks at the physical layer [8-14]. To explore the spatial diversity potential of the relaying networks and to boost the secrecy capacity (the difference between the channel capacity of the legitimate main link and that of the eavesdropping link), most of the existing work has been focused on secrecy-enhancing beamforming [8][9], as well as on intelligent relay node/jammer node (RN/JN) selection, etc.
Notably, given the availability of multiple relays, appropriately designed RN/JN selection is capable of achieving a significant security improvement for cooperative networks, which is emerging as a promising research topic. In particular, Zou \emph{et al.} investigated both amplify-and-forward (AF) as well as decode-and-forward (DF) based optimal relay selection conceived for enhancing the PLS in cooperative wireless networks [10] and [11], where the global channel state information (CSI) of both the main link and of the eavesdropping link was assumed to be available. Similarly, jamming techniques which impose artificial interference on the eavesdropper have also attracted substantial attention [12]-[14]. More specifically, several sophisticated joint relay and jammer selection schemes were proposed in [12], where the beneficially selected relay increases the reliability of the main link, while the carefully selected jammer imposes interference on the eavesdropper and simultaneously protects the legitimate destination from interference. In [13] and [14], cooperative jamming has been studied in the context of bidirectional scenarios and efficient RN/JN selection criteria have been developed for achieving improved secrecy rates with the aid of multiple relays. Furthermore, more effective relaying and jamming schemes when taking the information leakage of the source-eavesdropper link into consideration have been presented lately in [15] and [16].

Nevertheless, an idealized assumption of the previously reported research on PLS is the availability of perfect CSI, which is regarded as a stumbling block in the way of invoking practical secrecy-enhancing Wyner coding, on-off design, beamforming/precoding, as well as RN/JN selection. However, this idealized simplifying assumption is not realistic, since practical channel estimation (CE) imposes CSI imperfections, which are aggravated by the feedback delay, limited-rate feedback and channel estimation errors (CEE), etc [17].
Generally, the related research has been focused on the issues of robust secure beamforming design from an average secrecy rate based optimization perspective for point-to-point multi-antenna aided channels and relay channels [18][19] supporting delay-tolerant systems. For systems imposing stringent delay constraints, especially in imperfect CSI scenarios, perfect secrecy cannot always be achieved. Hence, the secrecy outage-based characterization of systems is more appropriate, which provides a probabilistic performance measure of secure communication. The concept of secrecy-outage was adopted in [20] for characterizing the probability of having both reliable and secure transmission, which, however, is inapplicable for the imperfect CSI case and fails to distinguish a connection-outage from the secrecy-outage. [21] proposed an alternative secrecy-outage formulation for characterizing the attainable security level and provided a general framework for designing transmission schemes that meet specific target security requirements. In order to quantify both the reliability and security performance at both the legitimate and the eavesdropper nodes separately, two types of outages, namely the connection outage probability (COP) and secrecy outage probability (SOP) are introduced. Then, considering the impact of time delay causing by the antenna selection process at the legitimate receiver, Hu \emph{et. al} in [22] proposed a new secure transmission scheme in the multi-input multi-output multi-eavesdropper wiretap channel. Much recently, considering the outdated CSI from the legitimate receiver, a new secure on-off transmission scheme was proposed for enhancing the secrecy throughput in [23].

Moreover, prior studies of the outage-based secure transmission design are limited to single-antenna assisted single-hop systems and have not been considered for cooperative relaying systems. Hence, the issues of secure transmissions over cooperative relaying channels expressed in terms of the SOP, COP and secrecy throughput constitute an open problem. On the other hand, apart from CEE, the CSI feedback delay results in critical challenges for the PLS of cooperative relaying systems, especially, when considering the specifics of RN/JN selection. The authors of [15] have also investigated the effects of outdated CSI knowledge concerning the legitimate links on the ergodic secrecy rate achieved by the proposed secure transmission
strategy in the context of DF relaying. The impact of CSI feedback delay on the secure relay and jammer selection conceived for DF relaying was investigated in [24], albeit only in terms of the SOP. In our previous study [25], we considered the secure transmission design and secrecy performance of an opportunistic DF system relying on outdated CSI, where only a single relay is invoked. Additionally, during the revision of this work, we investigated the security performance for outdated AF relay selection in [26]. Therefore, in this treatise, we extend our investigations to the PLS of multiple AF relaying assisted networks relying on RN/JN selection.

Explicitly, we focus our attention on the outage-based characterization of secure transmissions in cooperative relay-aided networks relying on realistic CSI feedback delay. To exploit the multi-relay induced diversity gain and the associated jamming capabilities, joint AF relay node and jammer node selection is employed by the relay-destination link. We assume that in line with the practical reality, the instantaneous eavesdropper's CSI is unavailable at the legitimate transmitter and that the RN/JN selections are performed based on the outdated CSI of the main links. Two types of cooperative strategies are invoked by our cooperative network operating under secrecy constraints, namely the traditionally best relay selection (TBRS) strategy as well as the joint relay and jammer selection (JRJS) strategy. Specifically, the main contributions of this paper can be summarized as follows:
\begin{itemize}
\item We develop an outage-based characterization for quantifying both the reliability and security performance of a two-hop AF relaying system. Specifically, in contrast to [21][22], we propose the novel definition of the reliable-and-secure connection probability (RSCP). Explicitly, closed-form expressions of the COP, SOP, and RSCP are derived for both the TBRS and for our JRJS strategies. Numerical results demonstrate that the JRJS scheme outperforms the TBRS scheme in terms of its RSCP.
\item We also introduce the reliability-security-ratio (RSR) for characterizing their direct relationship by a single parameter through the asymptotic analysis of the COP and SOP in the high-SNR regime. We derive the RSR for both the TBRS and JRJS strategies for investigating the effect of secrecy codeword rate setting, as well as that of the feedback delay and that of the power sharing ratio between the relay and the jammer on the RSR.
\item We then modify the definition of effective secrecy throughput by multiplying the secrecy rate with the RSCP, which results in an optimization problem of the transmit SNR, secrecy codeword rate and power sharing between the relay and jammer. It is shown that compared to the TBRS strategy, the JRJS achieves a significantly higher effective secrecy throughput, and the corresponding throughput loss is more sensitive to feedback delays. The impact of the direct source-eavesdropper link and additional throughput performance comparisons with respect to other related selection schemes are further discussed.
\end{itemize}

The remainder of this paper is organized as follows. Section II introduces our system model and describes both the TBRS and our JRJS strategies. In Section III and Section IV, we present the mathematical framework of our performance analysis both for the TBRS strategy and for the JRJS strategy, respectively, including the COP, SOP, RSCP, RSR and the effective secrecy throughput. Our numerical results and discussions are provided in Section V. Finally, Section VI offers our concluding remarks.

\section{SYSTEM MODEL}
\subsection{System Description}
\begin{figure}
\begin{center}
  \includegraphics[width=3.5in,angle=0]{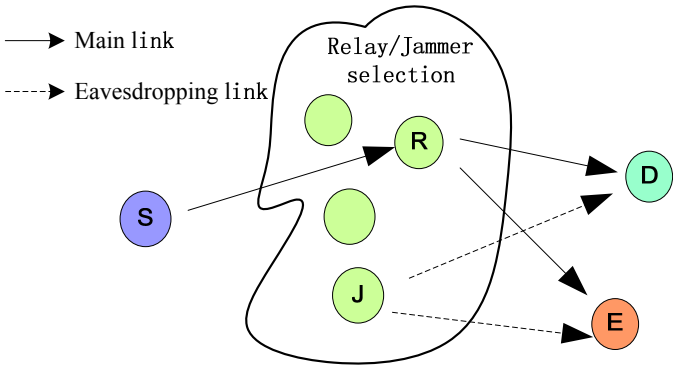}\\
  \caption{A cooperative relaying network assisted by multiple relays in the presence of an eavesdropper.}
\end{center}
\end{figure}

Consider a cooperative relaying network consisting of a source $S$, a destination $D$, $K_r$ relays $R_k$, $k = 1, \cdots ,K_r$, and an eavesdropper $E$, as shown in Fig. 1, where all nodes are equipped with a single transmit antenna (TA), except for the source which has $N_t$ TAs. The cooperative relay
architecture of Fig. 1 is generally applicable to diverse
practical wireless systems in the presence of an eavesdropper, including the family of wireless sensor networks (WSNs), mobile ad hoc networks (MANETs) and the long term evolution (LTE)-advanced cellular systems [11].

To exploit the diversity potential of multiple relay nodes over independently fading channels, AF relay/jammer selection is employed.
All relays operate in the half-duplex AF mode and data transmission is performed in two phases. More particularly, during the broadcast phase, the source node transmits its signal to a selected relay with the aid of beamforming (BF), which is invoked for forwarding the signal received from $S$ to $D$. An inherent assumption is that the transmit BF weights are based on the CSI estimates quantified and fed back by the selected relay.
During the cooperative phase, a pair of appropriately selected relays transmit towards $D$ and $E$, respectively. A conventional relay (denoted as $R^* $) forwards the source's message to the destination. Another relay (denoted as $J^* $) operates in the ``jammer mode" and imposes intentional interference upon $E$ in order confuse it. However, the $D$ is unable to mitigate the artificial interference emanating from the jammer node $J^* $ due to its critical secrecy constraints [12]. It should be noted that both the process of RN/JN selection and the feedback of the transmit BF weights from $R^* $ to the $S$ may impose a time-lag between the data transmission and channel estimation. These time delays are denoted by $T_{d_{SR} } $ and $T_{d_{RD} } $, respectively. Furthermore, we assume that the BF and RN/JN selection process is based on the perfectly estimated but outdated CSI. We employ the first-order autoregressive outdated CSI model of [20], whilst relying on the correlation coefficients of $\rho _{SR}  = J_0 (2\pi f_d T_{d_{SR} } )$ and $\rho _{RD}  = J_0 (2\pi f_d T_{d_{RD} } )$ for the two hops, where $J_0 \left(  \cdot  \right)$ is the zero-order Bessel function of the first kind and $f_d $ is the Doppler frequency.

A slow, flat, block Rayleigh fading environment is assumed, where the channel remains static for the coherence interval (one slot) and changes independently in different coherence intervals, as denoted by $h_{i,j}  \sim {\cal C}{\cal N}(0,\sigma _{i,j}^2 )
$, $i,j \in \left\{ {S,R,J,D,E} \right\}$. The  direct communication links are assumed to be unavailable due to the presence of obstructions between $S$ and $D$ as well as the eavesdropper\footnote{The case when the $S \to E$ link is introduced will be investigated separately in Section VI.}. This assumption
follows the rationale of [12] and has been routinely exploited in previous literature (see [27][28] and the references therein), where the source and relays belong to the same cluster, while the
destination and the eavesdropper are located in another. More specifically, this assumption
is especially valid in networks with broadcast and unicast transmission,
where each terminal is a legitimate receiver for one
signal and acts as an eavesdropper for some other signal. Therefore, the security concerns are only related to the cooperative relay-aided channel. Furthermore, additive white Gaussian noise (AWGN) is assumed with zero mean and unit variance $N_0$. Let $P_i$ be the transmit power of node $i$ and the instantaneous signal-to-noise ratio (SNR) of the $i \to j$ link is given by $\gamma _{i,j}  = {{P_i \left| {h_{i,j} } \right|^2 } \mathord{\left/
 {\vphantom {{P_i \left| {h_{i,j} } \right|^2 } {N_0 }}} \right.
 \kern-\nulldelimiterspace} {N_0 }}$.

We employ the constant-rate Wyner coding scheme for constructing wiretap codes of [2] in order to meet the PLS requirements due to the fact that the accurate global CSI is not available. Let $\mathbb{C}\left( {R_0 ,R_s ,N} \right)$ denote the set of all possible Wyner codes of length $N$, where $R_0 $ is the codeword transmission rate and $R_s $ is the confidential information rate ($R_0  > R_s $). The positive rate difference $R_e  = R_0  - R_s $ is the cost of providing secrecy against the eavesdropper. A confidential message  is encoded into a codeword at $S$ and then transmitted to $D$.
\subsection{Secure Transmission}
In the broadcast phase, $S$ transmits its BF signal $s(t)$ to the selected relay $R^* $, where the relay selection is performed before data transmission commences and the selection criterion will be detailed later in the context of the cooperative phase. The transmit BF vector ${\bf{w}}(\left. t \right|T_d )$ is calculated using the perfectly estimated but outdated CSI given by ${\bf{w}}(\left. t \right|T_{d_{SR} } ) = {{{\bf{h}}_{SR^* }^H (t - T_d )} \mathord{\left/
 {\vphantom {{{\bf{h}}_{SR^* }^H (t - T_d )} {\left| {{\bf{h}}_{SR^* } (t - T_{d_{SR} } )} \right|}}} \right.
 \kern-\nulldelimiterspace} {\left| {{\bf{h}}_{SR^* } (t - T_{d_{SR} } )} \right|}}$ [29], where we have ${\bf{h}}_{SR^* } (t) = \left[ {h_{SR^* ,1} (t), \ldots ,h_{SR^* ,N_t } (t)} \right]^T $, and the signal received by the relay $R^* $ can be written as
\begin{equation}\label{1}
y_{R^* } (t) = \sqrt {P_s } {\bf{w}}(\left. t \right|T_d ){\bf{h}}_{SR^* } (t)s(t) + n_{SR^* } (t),
\end{equation}
where $n_{SR^* } (t)$ is the AWGN at the relay. Then we can define the received SNR at the relay node as $\gamma _{SR}  = {{P_S \left| {{\bf{w}}(\left. t \right|T_{d_{SR} } ){\bf{h}}_{SR^* } (t)} \right|^2 } \mathord{\left/
 {\vphantom {{P_S \left| {{\bf{w}}(\left. t \right|T_{d_{SR} } ){\bf{h}}_{SR^* } (t)} \right|^2 } {N_0 }}} \right.
 \kern-\nulldelimiterspace} {N_0 }}$.

In the cooperative phase, we consider two RN/JN selection schemes performed by $D$: relay selection without jamming, as well as joint relay and jammer selection, respectively.

\subsubsection{Traditional Best Relay Selection}
The first category of solutions does not involve a jamming process and therefore only a conventional relay accesses the channel during the second phase of the protocol. The relay selection process is performed based on the highest instantaneous SNR of the second hop, formulated as
\begin{equation}\label{2}
R^*  = \arg \mathop {\max }\limits_{R_k  \in {\cal R}} \left\{ {\frac{{\tilde \gamma _{R_k D} }}{{\mathbb{E}\left[ {\gamma _{R_k E} } \right]}} = \frac{{P_R \left| {\tilde h_{R_k D} (t - T_d )} \right|^2 }}{{N_0 \mathbb{E}\left[ {\gamma _{R_k E} } \right]}}} \right\},
\end{equation}
where $\tilde \gamma _{R_k D} $ is the instantaneous SNR in the relay selection process, $\mathbb{E}\left[ {\gamma _{R_k E} } \right]$ denotes the average SNR at $E$.
We can model $\gamma _{R_k D} $ and $\tilde \gamma _{R_k D}$ as two gamma distributed RVs having the correlation factor of $\rho _{RD}^2 $.

During the second phase, the received signal $y_{R^* } (t)$ is multiplied by a time-variant AF-relay gain $G$ and retransmitted to $D$, where we have $G = \sqrt {{{P_R } \mathord{\left/
 {\vphantom {{P_R } {\left( {P_S \left| {{\bf{w}}_{opt} (\left. t \right|T_{d_{SR} } ){\bf{h}}_{SR^* } (t)} \right|^2  + N_0 } \right)}}} \right.
 \kern-\nulldelimiterspace} {\left( {P_S \left| {{\bf{w}}_{opt} (\left. t \right|T_{d_{SR} } ){\bf{h}}_{SR^* } (t)} \right|^2  + N_0 } \right)}}} $. After further mathematical manipulations, the mutual information (MI) between $S$, and $D$ as well as the eavesdropper can be written as
\begin{equation}\label{3}
I_D^{TBRS}  \!\!=\!\! \frac{1}{2}\log \left( {1 \!\!+\!\! \gamma _D^{TBRS} } \right) \!\!=\!\! \frac{1}{2}\log \left( {1 \!\!+\!\! \frac{{\gamma _{SR} \gamma _{R^* D} }}{{\gamma _{SR}  \!+\! \gamma _{R^* D}  \!+\! 1}}} \right)
\end{equation}
and
\begin{equation}\label{4}
I_E^{TBRS}  \!\!=\!\! \frac{1}{2}\log \left( {1 \!\!+\!\! \gamma _E^{TBRS} } \right) \!\!=\!\! \frac{1}{2}\log \left( {1 \!\!+\!\! \frac{{\gamma _{SR} \gamma _{R^* E} }}{{\gamma _{SR}  \!+\! \gamma _{R^* E}  \!+\! 1}}} \right).
\end{equation}

\subsubsection{Joint Relay and Jammer Selection}
Similarly, considering the unavailability of the instantaneous CSI regarding the eavesdropper, we adopt a suboptimal RN/JN selection metric conditioned on the outdated CSI as
\begin{equation}\label{5}
\begin{array}{l}
 R^*  = \arg \mathop {\max }\limits_{R_k  \in {\cal R}} \left\{ {\frac{{\tilde \gamma _{R_k D} }}{{\mathbb{E} \left[ {\gamma _{R_k E} } \right]}}} \right\}, \\
 J^*  = \arg \mathop {\min }\limits_{R_k  \in {\cal R} - R^* } \left\{ {\frac{{\tilde \gamma _{R_k D} }}{{\mathbb{E} \left[ {\gamma _{R_k E} } \right]}}} \right\}, \\
 \end{array}
\end{equation}
where $J^* $ is selected for minimizing the interference imposed on $D$.

It should be noted that to having the same transmit power as that of the TBRS case, we assume $P_{R*}  + P_{J^* }  = P_R $ for our JRJS strategy and introduce $\lambda  = {{P_{R^* } } \mathord{\left/
 {\vphantom {{P_{R^* } } {\left( {P_{R^* }  + P_{J^* } } \right)}}} \right.
 \kern-\nulldelimiterspace} {\left( {P_{R^* }  + P_{J^* } } \right)}}$ as the ratio of the relay's transmit power to the total power required by the active relay and jammer.

In the cooperative phase, $R^* $ will also amplify the received signal $y_{R^* } (t)$ by $G$ and forward it to $D$. At the same time, the jammer $J^* $ will generate intentional interference in order to confuse $E$, which will also cause interference at $D$. Consequently, the MI between the terminals is given by
\begin{equation}\label{6}
I_D^{JRJS}  \!\!=\!\! \frac{1}{2}\log \left( {1 \!\!+\!\! \gamma _D^{JRJS} } \right) \!\!=\!\! \frac{1}{2}\log \left( {1 \!\!+\!\! \frac{{\gamma _{SR} \frac{{\gamma _{R^* D} }}{{\gamma _{J^* D}  + 1}}}}{{\gamma _{SR}  \!\!+\!\! \frac{{\gamma _{R^* D} }}{{\gamma _{J^* D}  + 1}} \!\!+\!\! 1}}} \right),
\end{equation}
\begin{equation}\label{7}
I_E^{JRJS}  \!\!=\!\! \frac{1}{2}\log \left( {1 \!\!+\!\! \gamma _E^{JRJS} } \right) \!\!=\!\! \frac{1}{2}\log \left( {1 \!\!+\!\! \frac{{\gamma _{SR} \frac{{\gamma _{RE} }}{{\gamma _{JE}  + 1}}}}{{\gamma _{SR}  \!\!+\!\! \frac{{\gamma _{RE} }}{{\gamma _{JE}  + 1}} \!\!+\!\! 1}}} \right).
\end{equation}

\begin{remark}\label{Remark:1}
Generally, the optimal RN/JN selection scheme should take into account the global SNR knowledge set  $\left\{ {\gamma _{SR} ,\gamma _{RD} ,\gamma _{RE} } \right\}$. However, given the potentially excessive implementational complexity overhead of the optimal selection schemes and the unavailability of the global CSI, we employ suboptimal selection schemes as in [12]\footnote{To further alleviate the cooperation related overhead, the selection criterion is based on the $R \to D$ link, since the second-hop plays a dominant role in determining the received SNR, because the first hop corresponds to a MISO channel with the aid of with multiple antennas and hence it is more likely to be better than the second hop. The optimal selection based on both hops is beyond the scope of this work.}. Furthermore, it is commonly assumed that the average SNR of the eavesdropper is available at the transmitter, which seem some how not reasonable. However, as stated in most of the literature, such as [12-22, 24-28, 30], provided
that the eavesdropper belongs to the network, which is also the case in our paper, the related assumption might still be deemed
reasonably. Additionally, as in [8, 11, 12, 24], for mathematical convenience we assume that the relaying channels are independent and identically distributed and that we have $\mathbb{E}\left[ {\gamma _{SR_k } } \right] = \bar \gamma _{SR} $, $\mathbb{E}\left[ {\gamma _{R_k D} } \right] = \bar \gamma _{RD} $ and $\mathbb{E}\left[ {\gamma _{R_k E} } \right] = \bar \gamma _{RE} $. The distances between the relays are assumed to be much smaller than the distances between relays and the
source/destination/evesdropper, hence the corresponding path losses among the different relays are approximately the same. This assumption is reasonable both for WSNs and for MANETs associated with a symmetric clustered relay configuration and it may also be satisfied is also valid by classic cellular systems in a statistical sense [11].
\end{remark}

\section{Secure Transmission without Jamming}
In this section, we endeavor to characterize both the reliability and security performance comprehensively of the TBRS scheme. We first derive closed-form expressions for both the COP and SOP. Then, the RSR is introduced through the asymptotic analysis of the COP and SOP. Furthermore, we propose the novel definition of the RSCP and the effective secrecy throughput.

\subsection{COP and SOP}
When the perfect instantaneous CSI of the eavesdropper's channel and even the legitimate users' channel is unavailable, alternative definitions of the outage probability may be adopted for the statistical characterization of the attainable secrecy performance, especially for delay-limited applications. Based on [31, Definition 2], perfect secrecy cannot be achieved, when we have $R_e  < I_E $, where $I_E $ denotes the MI between the source and eavesdropper. Encountering this event is termed as a secrecy outage. Furthermore, the destination is unable to flawlessly decode the received code words when $R_0  > I_D $, which is termed as a connection outage.
The grade of reliability and the grade of security maintained by a transmission scheme may then be quantified by the COP and SOP, respectively.

We continue by presenting our preliminary results versus the point-to-point SNRs. Let us denote the cumulative distribution function (CDF) and probability density function (PDF) of a random variable $X$ by $F_X (x)$ and $f_X (x)$, respectively. On one hand, the PDF of $\gamma _{SR} $ using [29, Eq. (15)], is given by
\begin{equation}\label{8}
f_{\!\gamma _{SR }\!} \!\left(\! x \!\right)\! \!=\! \sum\limits_{n = 0}^{N_t  \!-\! 1} {\!\!\!\left(\!\!\! {\begin{array}{*{20}c}
   {N_t  \!\!-\!\! 1}  \\
   n  \\
\end{array}} \!\!\!\right)\!\!\!\frac{{\rho _{SR}^{2(N_t  \!-\! 1 \!-\! n)} (\bar \gamma _{SR} (1 \!\!-\!\! \rho _{SR}^2 ))^n }}{{\bar \gamma _{SR}^{N_t } (N_t  \!-\! 1 \!-\! n)!}}x^{N_t  \!-\! 1 \!-\! n} e^{\frac{{ - x}}{{\bar \gamma _{SR} }}} },
\end{equation}
while its CDF is given by
\begin{equation}\label{9}
F_{\!\gamma _{SR}\! } \!\left( \!x \!\right)\! \!=\! 1 \!-\! \sum\limits_{n \!= \!0}^{N_t  \!- \!1} {\sum\limits_{m \!= \!0}^{N_t  \!-\! 1 \!-\! n} {\!\!\!\left(\!\!\! {\begin{array}{*{20}c}
   {N_t  \!\!-\!\! 1}  \\
   n  \\
\end{array}} \!\!\!\right)\!\!\!\frac{{\rho _{SR}^{2(N_t \! -\! 1 \!-\! n)} (1 \!\!- \!\!\rho _{SR}^2 )^n }}{{m!\bar \gamma _{SR}^m }}x^m e^{\frac{{ - x}}{{\bar \gamma _{SR} }}} } }.
\end{equation}

On the other hand, for the instantaneous SNR of the $R \to D$ hop, according to the principles of concomitants or induced order statistics, the CDF of $\gamma _{R^* D} $, can be derived as in [32]
\begin{equation}\label{10}
\!\!F_{\gamma _{R^*D} } \left( y \right)\!\! \!=\! K_r \sum\limits_{k = 0}^{K_r  - 1} {( \!-\! 1)^k \!\!\!\left(\!\!\! {\begin{array}{*{20}c}
   {K_r  - 1}  \\
   k  \\
\end{array}} \!\!\!\right)\!\!\!\frac{{1 \!\!-\!\! e^{\frac{{ - (k + 1)y}}{{(k(1 \!-\! \rho _{RD}^2 ) \!+\! 1)\bar \gamma _{RD} }}} }}{{k + 1}}}.
\end{equation}

Thus, the COP of the TBRS strategy is given by
\begin{equation}\label{11}
 P_{co}^{TBRS}\left( R_0 \right)  = \Pr \left[ {I_D^{TBRS}  < R_0 } \right] = F_{\gamma _D^{TBRS} } \left( {\gamma _{th}^D } \right) \\,
\end{equation}
where we have $ \gamma _{th}^D  = 2^{2R_0 }  - 1$ and the CDF of $\gamma _D^{TBRS} $ can be calculated as
\begin{equation}\label{12}
F_{\gamma_D^{TBRS} } \left( x \right) \!=\! 1 \!\!-\!\! \int_0^\infty  {\!\left[\! {1 \!\!-\!\! F_{\gamma _{R^*D} } \!\!\left(\!\! {\frac{{xz \!+\! x(x \!+\! 1)}}{z}} \!\right)\!} \!\right]\!f_{\gamma _{SR^* } } \!\left(\! {z \!\!+\!\! x} \!\right)\!dz}.
\end{equation}

Consequently, by substituting (8) and (10) into (12), and using [33, Eq. (3.471.9)], we arrive at a closed-form expression for $F_{\gamma _D^{TBRS} } \left( x \right)$ as
\begin{equation}\label{13}
\begin{array}{l}
 F_{\gamma _D^{TBRS} } \left( x \right) \!=\! 1 \!\!-\!\! 2\sum\limits_{n = 0}^{N_t  \!-\! 1} {\sum\limits_{k = 0}^{K_r  \!-\! 1} {\sum\limits_{m = 0}^{N_t  \!-\! 1 \!-\! n} {( \!-\! 1)^k } } } K_r \!\!\!\left(\!\!\! {\begin{array}{*{20}c}
   {N_t  \!\!-\!\! 1}  \\
   n  \\
\end{array}} \!\!\right)\!\!\!\!\left(\!\! {\begin{array}{*{20}c}
   {K_r  \!\!-\!\! 1}  \\
   k  \\
\end{array}} \!\!\!\!\right)\!\!\!\! \\
 \times \!\!\!\left(\!\!\! {\begin{array}{*{20}c}
   {N_t  \!\!-\!\! 1 \!\!-\!\! n}  \\
   m  \\
\end{array}} \!\!\!\right)\!\!\!\frac{{\rho _{SR}^{2(N_t  \!-\! 1 \!-\! n)} (1 \!-\! \rho _{SR}^2 )^n x^{N_t  \!-\! 1 \!-\! n \!-\! m} }}{{(N_t  \!-\! 1 \!-\! n)!(k \!+\! 1)\bar \gamma _{SR}^{N_t  \!-\! n} }}
 \left[ {\frac{{\bar \gamma _{SR} x\left( {x + 1} \right)}}{{\omega _k \bar \gamma _{RD} }}} \right]^{\frac{{m + 1}}{2}}\\
 \times e^{ - \left( {\frac{{\bar \gamma _{SR}  + \omega _k \bar \gamma _{RD} }}{{\omega _k \bar \gamma _{SR} \bar \gamma _{RD} }}} \right)x} K_{m + 1} \left( {2\sqrt {\frac{{x\left( {x \!+\! 1} \right)}}{{\omega _k \bar \gamma _{SR} \bar \gamma _{RD} }}} } \right){\rm{ }} \\
 \end{array},
\end{equation}
where we have $\omega _k  = \frac{{k(1 - \rho _{RD}^2 ) + 1}}{{k + 1}}$. Then, by substituting $x = \gamma _{th}^D $ into (13), we obtain $P_{co}^{TBRS} $.

Furthermore, the SOP of the TBRS strategy may be expressed as
\begin{equation}\label{14}
 P_{so}^{TBRS} \left( {R_0 ,R_s } \right) \!=\! \Pr \left[ {I_E^{TBRS}  \!>\! R_0  \!-\! R_s } \right] \! =\! 1 \!-\! F_{\gamma _E^{TBRS} } \left( {\gamma _{th}^E } \right),
\end{equation}
where we have $ \gamma _{th}^E  = 2^{2\left( {R_0  - R_s } \right)}  - 1$. Similarly, we may calculate the CDF of $\gamma _E^{TBRS} $ in (14) as
\begin{equation}\label{15}
\begin{array}{l}
 F_{\gamma _E^{TBRS} } \left( x \right) \!=\! 1 \!\!-\!\! 2\sum\limits_{n = 0}^{N_t  \!-\! 1} {\sum\limits_{m = 0}^{N_t  \!-\! 1 \!-\! n} { \!\!\!\left(\!\!\! {\begin{array}{*{20}c}
   {N_t  \!\!-\!\! 1}  \\
   n  \\
\end{array}} \!\!\right)\!\!} }\!\!\left(\!\! {\begin{array}{*{20}c}
   {N_t  \!\!-\!\! 1 \!\!-\!\! n}  \\
   m  \\
\end{array}} \!\!\!\right)\!\!\! \\
 {\rm{           }} \times \frac{{\rho _{SR}^{2(N_t  - 1 - n)} (1 - \rho _{SR}^2 )^n x^{N_t  - 1 - n - m} }}{{(N_t  - 1 - n)!\bar \gamma _{SR}^{N_t  - n} }}\left[ {\frac{{\bar \gamma _{SR} x\left( {x + 1} \right)}}{{\bar \gamma _{RE} }}} \right]^{\frac{{m + 1}}{2}}  \\
 {\rm{           }} \times e^{ - \left( {\frac{{\bar \gamma _{SR}  + \bar \gamma _{RE} }}{{\bar \gamma _{SR} \bar \gamma _{RE} }}} \right)x} K_{m + 1} \left( {2\sqrt {\frac{{x\left( {x + 1} \right)}}{{\bar \gamma _{SR} \bar \gamma _{RE} }}} } \right){\rm{ }} \\
 \end{array}.
\end{equation}

Then, by substituting $x = \gamma _{th}^E $ into (15), we can derive $P_{so}^{TBRS} $.

The COP and SOP in (11) and (14) characterize the attainable reliability and security performance, respectively, and can be regarded as the detailed requirements of accurate system design. From the definition of COP and SOP, it is clear that the reliability of the main link can be improved by increasing the transmit SNR (or decreasing its data rate) to reduce the COP, which unfortunately increases the risk of eavesdropping. Thus, a tradeoff between reliability and security may be struck, despite the fact that closed-from expressions cannot be obtained as in [11]. Furthermore, we denote the minimal reliability and security requirements by $\upsilon $ and $\delta $, where the feasible range of the reliability constraint is $0 < \upsilon  < 1$. Bearing in mind that the COP is a monotonously increasing function of $R_0 $, the corresponding threshold of the codeword transmission rate is $R_0^{th}  = \arg \left\{ {P_{co}^{TBRS} \left( {R_0 } \right) = \upsilon } \right\}$, which leads to a lower bound of the SOP, when we have $\left({R_0  - R_s} \right) \to R_0^{th} $. Thus, the feasible range of $\delta $ is $P_{so}^{TBRS} \left( {R_0^{th} ,0} \right) < \delta  < 1$. The above analysis indicates that given a reliability constraint $\upsilon $, the lower bound of the security constraint is determined.

\subsection{Reliability-Security Ratio}
In this subsection, we will focus our attention on the asymptotic analysis of the COP and SOP in the high-SNR regime. Then, inspired by [25], we introduce the concept of the reliability-security ratio (RSR) for characterizing the direct relationship between reliability and security.

\textbf{Proposition 1:}
Based on the asymptotic probabilities of $P_{co} $ and $P_{so} $ at high SNRs\footnote{Assuming equal power allocation between $S$ and the relay, yielding $P_S  = P_R  = P$, and define $\eta  = {P \mathord{\left/
 {\vphantom {P {N_0 }}} \right.
 \kern-\nulldelimiterspace} {N_0 }}$ as the transmit SNR [24].}, the reliability-security ratio is defined as
\begin{equation}\label{16}
P_{co} \left( {R_0 } \right) = \Lambda \left[ {1 - P_{so} \left( {R_0 ,R_s } \right)} \right],
\end{equation}
where $\Lambda  = \lim _{\eta \to \infty } {{P_{co} } \mathord{\left/
 {\vphantom {{P_{co} } {\left( {1 - P_{so} } \right)}}} \right.
 \kern-\nulldelimiterspace} {\left( {1 - P_{so} } \right)}}$, which represents the improvement in COP upon decreasing the SOP. More specifically, since the reduction of the SOP/COP must be followed by an improvement of COP/SOP, a lower $\Lambda$ implies that when the security is reduced, the reliability is improved and vice versa.
 Thus, for the TBRS scheme studied above, the RSR is derived as
\setcounter{equation}{16}
\begin{figure*}[ht]
\begin{equation}\label{17}
\Lambda ^{TBRS}  = \frac{{\left[ {\left( {1 - \rho _{SR}^2 } \right)^{N_t  - 1}  + \sum\limits_{k = 0}^{K_r  - 1} {( - 1)^k \left( {\begin{array}{*{20}c}
   {K_r  - 1}  \\
   k  \\
\end{array}} \right)} \frac{{K_r \sigma _{SR}^2 }}{{\left[ {k\left( {1 - \rho _{RD}^2 } \right) + 1} \right]\sigma _{RD}^2 }}} \right]\left( {2^{2R_0 }  - 1} \right)}}{{\left[ {N_t \left( {1 - \rho _{SR}^2 } \right)^{N_t  - 1}  + {{\sigma _{SR}^2 } \mathord{\left/
 {\vphantom {{\sigma _{SR}^2 } {\sigma _{RE}^2 }}} \right.
 \kern-\nulldelimiterspace} {\sigma _{RE}^2 }}} \right]\left( {2^{2\left( {R_0  - R_s } \right)}  - 1} \right)}}
\end{equation}
\end{figure*}
\setcounter{equation}{17}

\emph{Proof:} The proof is given in Appendix B.

\begin{remark}\label{Remark:2}
It can be seen from the above expression that the factor $\Lambda $ is independent of the transmit SNR, but directly depends on the channel gains, the rate-pair $\left( {R_0 ,R_s } \right)$ and on the number of transmit antennas and relays. For a given $R_s $, reducing $R_0 $ to enhance the reliability may erode the security, because $\left(R_0  - R_s \right)$ is also reduced. Conversely, increasing $R_0 $ provides more redundancy for protecting the security of the information, but simultaneously the reliability is reduced. Hence, the RSR analysis underlines an important point of view concerning how to balance the reliability vs security trade-off by adjusting $\left( {R_0 ,R_s } \right)$. Furthermore, as long as a CSI feedback delay exists, the RSR has an intimate relationship with $\rho _{SR} $ and $\rho _{RD} $. It is clear that the value of $\Lambda ^{TBRS} $ decreases as $\rho _{RD} $ increases, which is due to the fact that the relay selection process only improves the reliability of the legitimate user. On the other hand, since we always have the conclusion that $\sum\limits_{k = 0}^{K_r  \!-\! 1} {( \!-\! 1)^k \!\!\left(\!\! {\begin{array}{*{20}c}
   {K_r  \!-\! 1}  \\
   k  \\
\end{array}} \!\!\right)\!\!} \frac{{K_r }}{{k\left( {1 \!-\! \rho _{RD}^2 } \right) \!+\! 1}} < 1$, when $\sigma _{RD}^2 $ and $\sigma _{RE}^2 $ are comparable, $\Lambda ^{TBRS} $ will be reduced as $\rho _{SR} $ increases. This observation implies that although both $P_{co} $ and $\left(1 - P_{so}\right)$ are reduced when the first hop CSI becomes better, the improvement of the reliability is more substantial than the security loss, as $\rho _{SR} $ increases.
\end{remark}
\subsection{Effective Secrecy Throughput}
It should be noted that the COP and SOP metrics ignore the correlation between these two outage events. More specifically, in contrast to the point-to-point transmission case, since the $S \to R$ link's SNR included in the MI expressions of (3) and (4), the secrecy outage and the connection outage are definitely not independent of each other. Therefore, it might be of limited benefit in evaluating the reliability or security separately. We note furthermore that although another metric referred to as the secrecy throughput was introduced as the product of the successful decoding probability and of the secrecy rate [21][22], this definition ignores the fact that a reliable transmission may be insecure and the SOP is not taken into consideration. Hence, this metric is unable to holistically characterize the efficiency of our scheme, while is capable of achieving both reliable and secure transmission. Therefore, in this section we redefine the effective secrecy throughput as the probability of a successful transmission (reliable and secure) multiplied by the secrecy rate, namely as $\varsigma  = R_s P_{R\& S} $, where the reliable-and-secure connection probability (RSCP) is defined as
\begin{equation}\label{18}
P_{R\& S}  = \Pr \left\{ {I_D  > R_0 ,I_E  < R_0  - R_s } \right\}.
\end{equation}

Upon substituting the expressions of $I_D$ and $I_E$ in (3) and (4) into (18), we can rewrite $P_{R\& S} $ for the TBRS strategy in (19).
\setcounter{equation}{18}
\begin{figure*}[ht]
\begin{equation}\label{19}
\begin{array}{l}
  P_{R\& S}^{TBRS}  \!=\! \Pr \left\{ {\left\{ {\gamma _{SR} \! >\! \gamma _{th}^D ,\gamma _{R^* D}  \!>\! \frac{{\gamma _{th}^D \gamma _{SR}  + \gamma _{SR} }}{{\gamma _{SR}  - \gamma _{th}^D }}} \right\} \cap \left[ {\left\{ {\gamma _{SR}  \!>\! \gamma _{th}^E ,\gamma _{R^* E}  \!<\! \frac{{\gamma _{th}^E \gamma _{SR}  + \gamma _{SR} }}{{\gamma _{SR}  - \gamma _{th}^E }}} \right\} \cup \left\{ {\gamma _{SR}  \!<\! \gamma _{th}^E } \right\}} \right]} \right\} \\
  ~~~~~~~~= \Pr \left\{ {\gamma _{SR}  > \gamma _{th}^D ,\gamma _{R^* D}  > \gamma _{th}^D  + \frac{{\gamma _{th}^D \left( {\gamma _{th}^D  + 1} \right)}}{{\gamma _{SR}  - \gamma _{th}^D }},\gamma _{R^* E}  < \gamma _{th}^E  + \frac{{\gamma _{th}^E \left( {\gamma _{th}^E  + 1} \right)}}{{\gamma _{SR}  - \gamma _{th}^E }}} \right\} \\
 \end{array}.
\end{equation}
\end{figure*}
\setcounter{equation}{19}

Finally, using the corresponding CDFs and PDFs of (8), (9) and (10) from our previous analysis, we can obtain $P_{R\& S}^{TBRS}$ in (20) as well as the secrecy throughput.
\setcounter{equation}{19}
\begin{figure*}[ht]
\begin{equation}\label{20}
\begin{array}{l}
 P_{R\& S}^{TBRS} = \int_{\gamma _{th}^D }^\infty  {\left[ {1 - F_{\gamma _{R^* D} } \left( {\gamma _{th}^D  + \frac{{\gamma _{th}^D \left( {\gamma _{th}^D  + 1} \right)}}{{x - \gamma _{th}^D }}} \right)} \right]F_{\gamma _{R^* E} } \left( {\gamma _{th}^E  + \frac{{\gamma _{th}^E \left( {\gamma _{th}^E  + 1} \right)}}{{x - \gamma _{th}^E }}} \right)f_{\gamma _{SR^* } } \left( x \right)dx}  \\
 \approx
2\sum\limits_{n = 0}^{N_t  \!-\! 1} {\sum\limits_{k = 0}^{K_r  \!-\! 1} {\sum\limits_{m = 0}^{N_t  \!-\! 1 \!-\! n} {( \!-\! 1)^k } \!\!\left(\!\! {\begin{array}{*{20}c}
   {K_r  \!\!-\!\! 1}  \\
   k  \\
\end{array}} \!\!\right)\!\!} \!\!\left(\!\! {\begin{array}{*{20}c}
   {N_t  \!\!-\!\! 1}  \\
   n  \\
\end{array}} \!\!\right)\!\!\!\!\left(\!\! {\begin{array}{*{20}c}
   {N_t  \!-\! 1 \!-\! n}  \\
   m  \\
\end{array}} \!\!\right)\!\!\frac{{K_r \rho _{SR}^{2(N_t  - 1 - n)} \left( {1 - \rho _{SR}^2 } \right)^n \left( {\gamma _{th}^D } \right)^{N_t  - 1 - n - m} }}{{\left( {N_t  - 1 - n} \right)!\left( {k + 1} \right)\bar \gamma _{SR}^{N_t  - n - {{\left( {m + 1} \right)} \mathord{\left/
 {\vphantom {{\left( {m + 1} \right)} 2}} \right.
 \kern-\nulldelimiterspace} 2}} }}} \\
 ~~~\times \exp \!\left[\! { - \!\!\left( \!\!{\frac{{\gamma _{th}^D }}{{\bar \gamma _{SR} }} \!+\! \frac{{\gamma _{th}^D }}{{\omega _k \bar \gamma _{RD} }}} \!\right)\!} \right]\!\left[\! {\!\left(\! {\frac{{\gamma _{th}^D \left( {\gamma _{th}^D  + 1} \!\right)\!}}{{\omega _k \bar \gamma _{RD} }}} \!\right)\!^{\frac{{m + 1}}{2}} K_{m \!+\! 1} \!\left(\! {2\sqrt {\frac{{\gamma _{th}^D \left( {\gamma _{th}^D  + 1} \right)}}{{\omega _k \bar \gamma _{SR} \bar \gamma _{RD} }}} } \right) }\right.\\
 ~~~\left. {\!-\! \exp \!\left(\! {\frac{{ - \gamma _{th}^E }}{{\bar \gamma _{RE} }}} \!\right)\!\!\left(\! {\frac{{\gamma _{th}^D \left( {\gamma _{th}^D  \!+\! 1} \right)}}{{\omega _k \bar \gamma _{RD} }} \!\!+\!\! \frac{{\gamma _{th}^E \left( {\gamma _{th}^E  \!+\! 1} \right)}}{{\bar \gamma _{RE}  \!+\! \gamma _{th}^D  \!-\! \gamma _{th}^E  }}} \!\right)\!^{\frac{{m + 1}}{2}} K_{m \!+\! 1} \!\left(\! {2\sqrt {\frac{{\gamma _{th}^D \left( {\gamma _{th}^D  + 1} \right)}}{{\omega _k \bar \gamma _{SR} \bar \gamma _{RD} }} \!\!+\!\! \frac{{\gamma _{th}^E \left( {\gamma _{th}^E  \!+\! 1} \right)}}{{\bar \gamma _{SR} \!\left(\! {\bar \gamma _{RE}  \!+\! \gamma _{th}^D  \!-\! \gamma _{th}^E } \!\right)\!}}} } \right)} \!\right]\! \\
 \end{array}.
\end{equation}
\end{figure*}
\setcounter{equation}{20}

Furthermore, considering the asymptotic result for RSCP at high SNRs in (20) by applying the approximation $K_v \left( x \right) \approx {{\left( {v - 1} \right)!} \mathord{\left/
 {\vphantom {{\left( {v - 1} \right)!} {2\left( {{x \mathord{\left/
 {\vphantom {x 2}} \right.
 \kern-\nulldelimiterspace} 2}} \right)^v }}} \right.
 \kern-\nulldelimiterspace} {2\left( {{x \mathord{\left/
 {\vphantom {x 2}} \right.
 \kern-\nulldelimiterspace} 2}} \right)^v }}$ and closing the highest terms of $\eta $ after invoking the McLaurin series representation for the exponential function, the asymptotic effective secrecy throughput can be approximated as in (21).
\setcounter{equation}{20}
\begin{figure*}[ht]
 \begin{equation}\label{21}
\tilde \varsigma ^{TBRS} \!\left(\! {R_0 ,R_s, \eta} \!\right)\! \!=\! R_s \!\!\left\{\!\! {1 \!\!-\!\! \!\left[\! {\frac{{N_t \!\left(\! {1 \!\!-\!\! \rho _{SR}^2 }\! \right)\!^{N_t  \!-\! 1} }}{{\sigma _{SR}^2 }} \!\!+\!\! \sum\limits_{k \!=\! 0}^{K_r  \!-\! 1} \frac{{K_r {( \!-\! 1)^k}}}{{\left[ {k\left( {1 \!-\! \rho _{RD}^2 } \right) \!+\! 1} \right]\sigma _{RD}^2 }}\!\!\!\left(\!\!\! {\begin{array}{*{20}c}
   {K_r  \!\!-\!\! 1}  \\
   k  \\
\end{array}} \!\!\right)\!\!} \!\right]\!\frac{{2^{2R_0 }  \!-\! 1}}{\eta }} \!\right\}\!\frac{{2^{2\left( {R_0  \!-\! R_s } \right)}  \!-\! 1}}{{\sigma _{RE}^2 \eta }}.
\end{equation}
\end{figure*}
\setcounter{equation}{21}

\begin{remark}\label{Remark:3}
Given the definition of COP, SOP and the secrecy throughput result of (21), it can be shown that for a fixed $R_s $, if $R_0 $ is too small, even though $P_{R\& S} $ may be high (i.e. close to one), the value of $\varsigma $ remains small. By contrast, if $R_0 $ is too large, the value of $P_{co} $ is close to one and therefore $\varsigma $ will also become small.  This observation is also suitable for $R_s $. Thus, as pointed out in the RSR analysis, it is elusive to improve both the reliability and security simultaneously, but both of them are equally crucial in terms of the effective secrecy throughput which depends on the rate pair $ \left( {R_0 ,R_s } \right)$.

Additionally, (21) also reveals that increasing the SNR would drastically reduce the effective secrecy throughput. For high transmit SNRs, a high reliability can indeed be perfectly guaranteed, but at the same time, the grade of the security is severely degraded. However, the probability of a reliable and simultaneously secure transmission will tend towards zero. Hence, we may conclude that there exists an optimal SNR which achieves the maximal secrecy throughput.

In conclusion, adopting the appropriate code rate pair and transmit SNR is crucial for achieving the maximum effective secrecy throughput, which can be formulated as
\begin{equation}\label{22}
\begin{array}{l}
 \mathop {\max }\limits_{R_0 ,R_s, \eta} {\rm{ }}\varsigma \left( {R_0 ,R_s } \right) = R_s P_{R\& S}^{TBRS}  \\
 s.t.{\rm{ }}P_{co}  \le \upsilon ,{\rm{  }}P_{so}  \le \delta ,{\rm{ }}0 < R_s  < R_0  \\
 \end{array},
\end{equation}
where $\upsilon $ and $\delta $ denote the system's reliability and security requirements. Unfortunately, it is quite a challenge to find the closed-form optimal solution to this problem due to the complexity of the expressions. Although suboptimal solutions can be found numerically (with the aid of gradient-based search techniques), the secrecy throughput optimization problem as well as the corresponding complexity analysis and performance comparisons are beyond the scop of this work.
\end{remark}

\section{Secure Transmission with Jamming}
In this section, we consider the extension of the above relay selection approaches to systems additionally invoking relay-aided jamming. The joint relay and jammer selection is based on the outdated but perfectly estimated CSI and the details have been presented in Section II. We would also like to investigate the security performance from an outage-based perspective. The COP, SOP, RSCP and effective secrecy throughput will be included.
\subsection{COP and SOP}
It is plausible that the main differences between the JRJS and TBRS schemes are determined by the instantaneous SNR of the $R \to D$ hop, where now a jammer is included. Based on our preliminary results detailed for the point-to-point SNRs in (8) and (10), we now focus our attention on the statistical analysis of the SNR including $J^* $. As stated for the JRJS scheme in Section II, $J^* $ corresponds to the lowest $\tilde \gamma _{R_k D} $ and is selected from the set $\left\{ {{\cal R} - R^* } \right\}$.  Recalling that $R^* $ is the best relay of the second hop, we have $\tilde \gamma _{J^* D}  = \min _{R_k  \in {\cal R} - R^* } \left\{ {\tilde \gamma _{R_k D} } \right\} \buildrel \Delta \over = \min _{R_k  \in {\cal R}} \left\{ {\tilde \gamma _{R_k D} } \right\}$ for $K_r  > 1$. Using the induced order statistics, the corresponding CDF of $\gamma _{R^* D} $ is presented in (10), while the PDF of $\gamma _{J^* D} $  can be formulated as
\begin{equation}\label{23}
f_{\gamma _{J^* D} } \left( x \right) = \frac{{K_r \exp \left( {\frac{{ - K_r x}}{{\left[ {\left( {K_r  - 1} \right)(1 - \rho _{RD}^2 ) + 1} \right]\bar \gamma _{JD} }}} \right)}}{{\left[ {\left( {K_r  - 1} \right)(1 - \rho _{RD}^2 ) + 1} \right]\bar \gamma _{JD} }}.
\end{equation}

Although the relay and jammer selection processes are not entirely disjoint, we may exploit the assumption that $\gamma _{R^* D} $ and $\gamma _{J^* D} $ are independent of each other, which is valid when the number of relays is sufficiently high, as justified in [24]. Let us define the signal to interference plus noise ratio (SINR) of the second hop as $\xi _D  = {{\gamma _{R^* D} } \mathord{\left/
 {\vphantom {{\gamma _{R^* D} } {\left( {\gamma _{J^* D}  + 1} \right)}}} \right.
 \kern-\nulldelimiterspace} {\left( {\gamma _{J^* D}  + 1} \right)}}$, using (10) and (23), whose CDF can be formulated as
\begin{equation}\label{24}
F_{\xi _D } \left( x \right) \!=\! 1 \!-\! K_r \sum\limits_{k = 0}^{K_r  \!-\! 1} {( \!-\! 1)^k \!\!\left(\!\! {\begin{array}{*{20}c}
   {K_r  \!\!-\!\! 1}  \\
   k  \\
\end{array}} \!\!\right)\!\!\frac{{\varphi _k }e^{\frac{{ - x}}{{\bar \gamma _{RD} \omega _k }}}}{{\left( {k \!+\! 1} \right)\left( {x \!+\! \varphi _k } \right)}}},
\end{equation}
where we have $\varphi _k  = \frac{{\lambda K_r \omega _k }}{{\left[ {\left( {K_r  - 1} \right)(1 - \rho _{RD}^2 ) + 1} \right]\left( {1 - \lambda } \right)}}$.

As far as the eavesdropper is concerned, $\gamma _{R^* E} $ and $\gamma _{J^* E} $ are independent and exponentially distributed. Furthermore, for $\xi _E  = {{\gamma _{R^* E} } \mathord{\left/
 {\vphantom {{\gamma _{R^* E} } {\left( {\gamma _{J^* E}  + 1} \right)}}} \right.
 \kern-\nulldelimiterspace} {\left( {\gamma _{J^* E}  + 1} \right)}}$, we have
\begin{equation}\label{25}
F_{\xi _E } \left( x \right) = 1 - \frac{\phi }{{x + \phi }}e^{\frac{{ - x}}{{\bar \gamma _{RE} }}},
\end{equation}
where $\phi  = {\lambda  \mathord{\left/
 {\vphantom {\lambda  {\left( {1 - \lambda } \right)}}} \right.
 \kern-\nulldelimiterspace} {\left( {1 - \lambda } \right)}}$. According to the definition of COP and SOP in the previous section, we can obtain the following closed-form approximations of the COP and SOP\footnote{When we have $\lambda \to 1$, (24) will degenerate into the TBRS case seen in (10). The performance analysis of the JRJS will be presented separately in the following, since several approximations have to be included.}.

\textbf{Lemma 1:} The COP and SOP of the JRJS strategy associated with feedback delays is approximated by
\begin{equation}\label{26}
\begin{array}{l}
 P_{co}^{JRJS} \!\left(\! {R_0 } \!\right)\!\!\! \approx\!\! 1 \!\!\!-\!\!\! \sum\limits_{n = 0}^{N_t  \!-\! 1} {\sum\limits_{k \!=\! 0}^{K_r\!-\!1 } {\sum\limits_{m = 0}^{N_t  \!-\! 1 \!-\! n} { \!\!\left(\!\! {\begin{array}{*{20}c}
   {N_t  \!\!-\!\! 1}  \\
   n  \\
\end{array}} \!\!\right)\!\!\!\!\left(\!\! {\begin{array}{*{20}c}
   {K_r\!\!-\!\!1 }  \\
   k  \\
\end{array}} \!\!\right)\!\!\!\!\left( \!\!{\begin{array}{*{20}c}
   {N_t  \!\!-\!\! 1 \!\!-\!\! n}  \\
   m  \\
\end{array}} \!\!\!\!\right)\!\!\!\!} } } \\
 \frac{( \!-\! 1)^k {\left( {K_r  + 1} \right)\rho _{SR}^{2(N_t  \!-\! 1 \!-\! n)} \left( {1 \!-\! \rho _{SR}^2 } \right)^n }}{{\left( {N_t  - 1 - n} \right)!\left( {k + 1} \right)\bar \gamma _{SR}^{N_t  - n} }}\frac{{\Gamma \left( {m \!+\! 2} \right)\hat \varphi _k \left( {\gamma _{th}^D } \right)^{N_t  - n} \left( {\gamma _{th}^D  + 1} \right)^{m + 1} }}{{\left( {\gamma _{th}^D  \!+\! \hat \varphi _k } \right)^{m \!+\! 2} }} \\
 \exp \left[ { - \frac{{\gamma _{th}^D \left( {\hat \varphi _k  - 1} \right)}}{{\bar \gamma _{SR} \left( {\gamma _{th}^D  + \hat \varphi _k } \right)}}} \right]\Gamma \left( { - m - 1,\frac{{\gamma _{th}^D (\gamma _{th}^D  + 1)}}{{\bar \gamma _{SR} \left( {\gamma _{th}^D  + \hat \varphi _k } \right)}}} \right) \\
 \end{array},
\end{equation}
where $\hat \varphi _k  = \frac{{K_r \lambda \omega _k \eta \sigma _{RD}^2 }}{{\left[ {\left( {K_r  - 1} \right)(1 - \rho _{RD}^2 ) + 1} \right]\left( {1 - \lambda } \right)\eta \sigma _{RD}^2  + K_r }}$,
and
\begin{equation}\label{27}
\begin{array}{l}
 P_{so}^{JRJS} \left( {R_0 ,R_s } \right) \!\approx\! \sum\limits_{n = 0}^{N_t  \!-\! 1} {\sum\limits_{m = 0}^{N_t  \!-\! 1 \!-\! n} {\!\!\left(\!\! {\begin{array}{*{20}c}
   {N_t  \!\!-\!\! 1}  \\
   n  \\
\end{array}} \!\!\right)\!\!\frac{{\rho _{SR}^{2(N_t  \!-\! 1 \!-\! n)} (1 \!-\! \rho _{SR}^2 )^n }}{{m!\bar \gamma _{SR}^m }}} }  \\
 ~~~~~~~~~~~~~~~~~~~~~~ \times \frac{{\left( {2\gamma _{th}^E } \right)^m \phi }}{{\left( {2\gamma _{th}^E  + \phi } \right)}}\exp \left[ { - \left( {\frac{{2\gamma _{th}^E }}{{\bar \gamma _{SR} }} + \frac{{2\gamma _{th}^E }}{{\bar \gamma _{RE} }}} \right)} \right] \\
 \end{array}.
\end{equation}
\emph{Proof:} The proof is given in Appendix B.

The feasible range of the reliability constraint is similar to that of the TBRS strategy and hence it is omitted here.
\subsection{Reliability-Security Ratio}
\textbf{Lemma 2:}
Recalling the definition in Section III, the RSR for the JRJS strategy may be expressed in (28).
\setcounter{equation}{27}
\begin{figure*}[ht]
 \begin{equation}\label{28}
\Lambda ^{JRJS}  \!\!=\!\! \frac{{\left( {2^{2R_0 }  \!-\! 1} \right)}}{{\!\left(\! {2^{2\left( {R_0  \!-\! R_s } \right)}  \!-\! 1} \!\right)\!}}\sum\limits_{k \!=\! 0}^{K_r  \!-\! 1} {\!\!\!\left(\!\!\! {\begin{array}{*{20}c}
   {K_r \!\! -\!\! 1}  \\
   k  \\
\end{array}} \!\!\!\right)\!\!\!} \frac{{( - 1)^k K_r \left[ {\left( {K_r  \!-\! 1} \right)(1 \!-\! \rho _{RD}^2 ) \!+\! 1} \right]\left[ {\left( {\lambda ^{ \!-\! 1}  \!-\! 1} \right)\left( {2^{2\left( {R_0  \!-\! R_s } \right)}  \!-\! 1} \right) \!+\! 1} \right]}}{{\left[ {\!\left(\! {K_r  \!-\! 1} \!\right)\!(1 \!\!-\!\! \rho _{RD}^2 ) \!\!+\!\! 1} \!\right]\!\left( {k \!+\! 1} \!\right)\!\left( {\lambda ^{ \!-\! 1}  \!-\! 1} \right)\left( {2^{2R_0 }  \!-\! 1} \right) \!+\! K_r \left[ {k(1 \!-\! \rho _{RD}^2 ) \!+\! 1} \!\right]\!}}.
\end{equation}
\end{figure*}
\setcounter{equation}{28}

\setcounter{equation}{30}
\begin{figure*}[ht]
 \begin{equation}\label{31}
\begin{array}{l}
 P_{R\& S}^{JRJS} \left( {R_0 ,R_s ,\lambda } \right) \!\approx\! \sum\limits_{n = 0}^{N_t  \!-\! 1} {\sum\limits_{k = 0}^{K_r  \!-\! 1} {\sum\limits_{m = 0}^{N_t  \!-\! 1 \!-\! n} {( \!-\! 1)^k } \!\!\left(\!\! {\begin{array}{*{20}c}
   {N_t  \!\!-\!\! 1}  \\
   n  \\
\end{array}} \!\!\right)\!\!\!\!\left(\!\! {\begin{array}{*{20}c}
   {K_r  \!\!-\!\! 1}  \\
   k  \\
\end{array}} \!\!\right)\!\!\!\!\left(\!\! {\begin{array}{*{20}c}
   {N_t  \!\!-\!\! 1 \!\!-\!\! n}  \\
   m  \\
\end{array}} \!\!\right)\!\!\frac{{K_r \rho _{SR}^{2(N_t  \!-\! 1 \!-\! n)} \left( {1 \!-\! \rho _{SR}^2 } \right)^n \hat \varphi _k \left( {\gamma _{th}^D } \right)^{N_t  - 1 - n - m} }}{{\left( {N_t  - 1 - n} \right)!\left( {k + 1} \right)\bar \gamma _{SR}^{N_t  - n} \left( {\gamma _{th}^D  + \hat \varphi _k } \right)e^{\frac{{\gamma _{th}^D }}{{\bar \gamma _{SR} }} + \frac{{\gamma _{th}^D }}{{\bar \gamma _{RD} \omega _k }}} }}} }  \\
 \!\left\{\! {\theta _{1,k}^{m + 1} e^{\frac{{\theta _1 }}{{\bar \gamma _{SR} }}} \Gamma \left( {m \!+\! 2} \right)\Gamma \!\!\left(\!\! { - m \!-\! 1,\frac{{\theta _{1,k} }}{{\bar \gamma _{SR} }}} \!\!\right)\!\! \!-\! \frac{{\hat \phi e^{ - {{\gamma _{th}^E } \mathord{\left/
 {\vphantom {{\gamma _{th}^E } {\bar \gamma _{RE} }}} \right.
 \kern-\nulldelimiterspace} {\bar \gamma _{RE} }}} }}{{\left( {\gamma _{th}^E  + \phi } \right)\left( {\theta _{1,k}  - \theta _2 } \right)}}} \right.\Gamma \left( {m \!+\! 3} \right)\left[ {\theta _2^{m + 2} e^{\frac{{\theta _2 }}{{\bar \gamma _{SR} }}} \Gamma \!\!\left(\!\! { - m \!-\! 2,\frac{{\theta _2 }}{{\bar \gamma _{SR} }}} \!\right)\! \!-\! \theta _{1,k}^{m + 2} e^{\frac{{\theta _1 }}{{\bar \gamma _{SR} }}} \Gamma \!\!\left(\!\! { - m \!-\! 2,\frac{{\theta _{1,k} }}{{\bar \gamma _{SR} }}} \!\right)\!} \right] \\
 +\left. {\Gamma \left( {m + 2} \right)\left( {\gamma _{th}^D  - \gamma _{th}^E } \right)\left[ {\theta _2^{m + 1} e^{\frac{{\theta _2 }}{{\bar \gamma _{SR} }}} \Gamma \left( { - m - 1,\frac{{\theta _2 }}{{\bar \gamma _{SR} }}} \right) - \theta _{1,k}^{m + 1} e^{\frac{{\theta _1 }}{{\bar \gamma _{SR} }}} \Gamma \left( { - m - 1,\frac{{\theta _{1,k} }}{{\bar \gamma _{SR} }}} \right)} \right]} \right\} \\
 \end{array}.
\end{equation}
\end{figure*}
\setcounter{equation}{28}

It can be seen from the above expression that in contrast to the analysis of the TBRS strategy operating without jamming, for a fixed SNR threshold, the CDF of the second-hop SNR will converge to a nonzero limit. We also find that this limit is determined by the power sharing ratio between the relay and the jammer. Furthermore, according to the analysis of the TBRS strategy, for $\eta  \to \infty $, we have $F_{\gamma _{SR^* } } \left( x \right) \to 0$. Thus, by exploiting the tight upper bound that  $\gamma _D^{TBRS}  \le \min \left\{ {\gamma _{SR} ,\gamma _{R^* D} } \right\}$ and $\gamma _E^{TBRS}  \le \min \left\{ {\gamma _{SR} ,\gamma _{R^* E} } \right\}$, we have $P_{co}^{JRJS,\infty }  \to F_{\gamma _{\xi _D } } \left( {\gamma _{th}^D } \right)$ and $1-P_{so}^{JRJS,\infty }  \to F_{\gamma _{\xi _E } } \left( {\gamma _{th}^E } \right)$. Finally, substituting the corresponding results into (16), we arrive at the RSR of the JRJS strategy.

\begin{remark}\label{Remark:4}
It can be seen from the RSR expression of (28) again that the rate-pair setting $\left( {R_0 ,R_s } \right)$ has an inconsistent influence on the RSR and hence we have to carefully adjust $R_0 $ and $R_s $ in order to balance the reliability versus security performance. Let us now focus our attention on the differences between the JRJS scheme and the TBRS arrangement.

Firstly, we may find that the power sharing ratio $\lambda $ between the relay and jammer plays a very important role. The optimization of $\lambda $ will be investigated from an effective secrecy throughput optimization point of view in the following subsection.

Secondly, it is plausible that in contrast to the behavior of the TBRS strategy, $\Lambda ^{JRJS}$  of (28) is only related to the delay of the second hop, but it is still a monotonically decreasing function of $\rho _{RD} $. This implies that the improvement of the channel quality of the JRJS will achieve a more pronounced COP improvement than the associated SOP improvement. Furthermore, recalling that the RSR is considered in the high-SNR region, it has no dependence on the first hop quality. This is due to the fact that if the first hop channel quality is sufficiently high for ensuring a successful transmission, the asymptotic CDFs of $\xi _D$ and $\xi _E$ in (29) and (30) associated with $\eta  \to \infty $ will converge to a nonzero limit at high SNRs, which ultimately dominates the COP and SOP.
\end{remark}
\subsection{Effective Secrecy Throughput}
Before proceeding to the effective secrecy throughput analysis, we also have to investigate the RSCP.

\textbf{Lemma 3:} The RSCP of our JRJS strategy may be approximated as in (31), where we have $\theta _{1,k}  = \frac{{\gamma _{th}^D \left( {\gamma _{th}^D  + 1} \right)}}{{\gamma _{th}^D  + \hat \varphi _k }}$, $\theta _2  = \gamma _{th}^D  - \gamma _{th}^E  + \frac{{\gamma _{th}^E \left( {\gamma _{th}^E  + 1} \right)}}{{\gamma _{th}^E  + \hat \phi }}$, and $\hat \phi  = \frac{{\lambda \eta \sigma _{RE}^2 }}{{\left( {1 - \lambda } \right)\eta \sigma _{RE}^2  + 1}}$.

\emph{Proof:} The proof is given in Appendix C.

Apart from the rate-pair $\left( {R_0 ,R_s } \right)$, the above-mentioned $P_{R\& S}^{JRJS}$ of (31) is also a function of the power sharing ratio $\lambda $ between the selected relay and the jammer.

Given the complexity of the RSCP expression, it is quite a challenge to find a closed-form result for the maximizing the effective secrecy throughput that $\mathop {\max }\limits_{{\rm{0 < }}\lambda  < 1}{\rm{ }}\varsigma  = R_s P_{R\& S}^{JRJS}$. Alternatively, we can focus on the asymptotic analysis in the high-SNR region and try to find a general closed-form solution for $\lambda$. Specifically, when we have $\eta  \to \infty $, $P_{R\& S}^{JRJS}$ will be dominated by the channel quality of the second hop, hence we have
\setcounter{equation}{31}
\begin{equation}\label{32}
\begin{array}{l}
 P_{R\& S}^{JRJS,\infty } \left( {R_0 ,R_s ,\lambda } \right) \approx \Pr \left\{ {\xi _D  > \gamma _{th}^D ,\xi _E  < \gamma _{th}^E } \right\} \\
  ~~~~~~~~~~~~~~~~~~~~~~~~~= \left[ {1 - F_{\xi _D } \left( {\gamma _{th}^D } \right)} \right]F_{\xi _E } \left( {\gamma _{th}^E } \right) \\
 \end{array},
\end{equation}
where the approximation is based on the fact that, in contrast to both $F_{\xi _D } \left( {\gamma _{th}^D } \right)$ and $F_{\xi _E } \left( {\gamma _{th}^E } \right)$, which converge to a nonzero limit regardless of $\eta $, the first hop's $F_{\gamma _{SR} } \left( x \right)$ will tend to zero and hence it can be neglected. Substituting the asymptotic results of (29) and (30) into (33), we can obtain $P_{R\& S}^{JRJS,\infty }$. In contrast to the TBRS case operating without jamming, as the SNR tends to $\infty $, the RSCP will tend to a nonzero value and upon increasing the transmit SNR beyond a certain limit will no longer increase the effective secrecy throughput.

Then, based on (32), we arrive at the approximated optimal value $\lambda _{opt} $, which is the solution of the following equation
\begin{equation}\label{33}
\frac{{\partial P_{R\& S}^{JRJS,\infty } \left( {R_0 ,R_s ,\lambda } \right)}}{{\partial \lambda }} = 0.
\end{equation}
Then, by exploiting the approximation of ${{\left[ {k(1 - \rho _{RD}^2 ) + 1} \right]} \mathord{\left/
 {\vphantom {{\left[ {k(1 - \rho _{RD}^2 ) + 1} \right]} {\left( {k + 1} \right)}}} \right.
 \kern-\nulldelimiterspace} {\left( {k + 1} \right)}} \approx 1 - \rho _{RD}^2$  in (29) for a large $\rho _{RD} $ (practically the CSI delay is small and $\rho _{RD}  \to 1$), we have
\begin{equation}\label{34}
\lambda _{subopt}  = \frac{{\sqrt {\left[ {\left( {K_r  - 1} \right)(1 - \rho _{RD}^2 ) + 1} \right]\gamma _{th} } }}{{\sqrt {\left[ {\left( {K_r  - 1} \right)(1 - \rho _{RD}^2 ) + 1} \right]\gamma _{th} }  + \sqrt {K_r (1 - \rho _{RD}^2 )} }},
\end{equation}
where $\gamma _{th}  = \left( {2^{2R_0 }  - 1} \right)\left( {2^{2\left( {R_0  - R_s } \right)}  - 1} \right)$. It is clearly this value is determined by the number of relays and $\left( {R_0 ,R_s } \right)$.


\section{Numerical Results}
Both our numerical and Monte-Carlo simulation results are presented in this section for verifying the theoretical PLS performance analysis of the multiple-relay aided network under CSI feedback delays. Explicitly, both the COP, SOP, RSCP, and RSR are validated for both the TBRS and JRJS strategies. Furthermore, the effect of feedback delays and system parameters (including the transmission rate pair $\left( {R_0 ,R_s } \right)$ and the power sharing ratio $\lambda $ between the relay and jammer) on the achievable effective secrecy throughput are evaluated. The Rayleigh fading model is employed for characterizing all communication links in our system. Additionally, we set the total power to $P = 1$, $\sigma _{SR}^2  = \sigma _{RD}^2  =\sigma _{RE}^2  = 1$, and used $T_{d_{SR} }  = T_{d_{RD} }  = T_d $.

Fig. 2 plots the COP and SOP versus the transmit SNR for both the TBRS and JRJS strategies in conjunction with different rate pairs. The analytical lines are plotted by using Eqs. (11) and (14) for the TBRS strategy, and by using Eqs. (26) and (27) for the TBRS case, respectively. It can be clearly seen from the figure that the analytical and simulated outage probability curves match well, which confirms the accuracy of the mathematical analysis. As expected, compared to the TBRS strategy, the SOP of the JRJS strategy is much better, while the COP is worse. We can also find that both the COP and SOP will converge to an outage floor at high SNRs for the JRJS strategy. The reason for this is that the jammer also imposes interference on the destination and the interference inflicted increases with the SNR. Thus, the designers have to take into account the tradeoff between the reliability as wel as security and the interference imposed on $D$, particularly when considering the JRJS strategy. Moreover, we can observe in Fig. 2 that increasing the transmission rate decreases the COP and increases the SOP.

\begin{figure}
\begin{center}
  \includegraphics[width=3in,height=3in,angle=0]{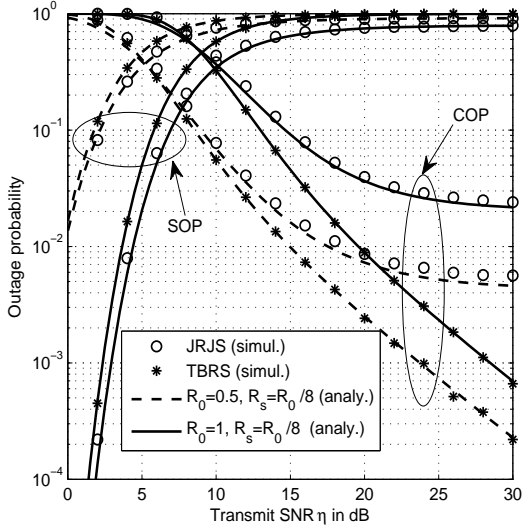}\\
  \caption{COP and SOP versus transmit SNR for the TBRS and JRJS strategies in conjunction with different rate pairs, for $N_t  = K_r  = 3$, $f_d T_d  = 0.1$, and $\lambda  = 1/10$.}
\end{center}
\end{figure}

Fig. 3 further characterizes the SOP versus COP for both the TBRS and JRJS strategies based on the numerical results in Fig. 2, which shows the tradeoff between the reliability and security. It can be seen from the figure that the SOP decreases as the COP increases, and for a specific COP, the SOP of the JRJS scheme is strictly lower than that of TBRS. This confirms that the JRJS scheme performs better than the conventional TBRS scheme. Furthermore, the CSI feedback delay will also degrade the system tradeoff performance.

\begin{figure}
\begin{center}
  \includegraphics[width=3in,height=3in,angle=0]{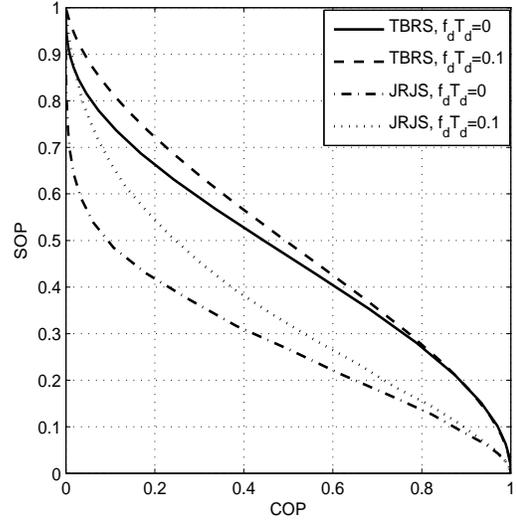}\\
  \caption{SOP versus COP for the TBRS and JRJS strategies with different feedback delays for $N_t  = K_r  = 3$, $R_s  = R_0 /8$, and $\lambda  = 1/10$.}
\end{center}
\end{figure}

\begin{figure}
\begin{center}
  \includegraphics[width=3in,height=3in,angle=0]{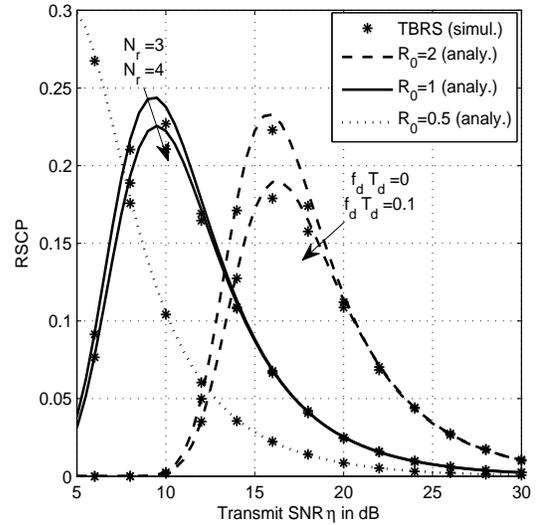}\\
  \caption{RSCP versus transmit SNR for the TBRS strategy with different rate pairs for $N_t  = K_r  = 3$, $f_d T_d  = 0.1$}
\end{center}
\end{figure}

Fig. 4 illustrates the RSCP versus transmit SNR for the TBRS strategy in the context of different network configurations, including different rate pairs, different number of relays, as well as both perfect and outdated CSI feedback scenarios. The analytical lines are plotted by using the approximation in (20). We may conclude from the figure that the rate pare setting $\left( {R_0 ,R_s } \right)$ determines both the reliability and security transmission performance. These curves also show that the RSCP is a concave function of the transmit SNR, while the continued boosting of the SNR would only decrease the probability of a successful transmission. We can observe from Fig.4 that for a high transmit SNR, total reliability can be guaranteed, whereas the associated grade of security is severely eroded. Furthermore, increasing the number of relays and decreasing the feedback delay will improve both the reliability and security performance

The RSCP of the JRJS strategy is presented in Fig. 5 for different power sharing ratios between relaying and jamming. Both the integration-form (45) and the approximated closed-form in (31) match well with the Monte-Carlo simulations. The performance of the TBRS strategy is also included for comparison. The JRJS scheme outperforms the TBRS operating without jamming under the scenario considered, when encountering comparable relay-destination and relay-eavesdropper channels. For some extreme configurations (when the relay-eavesdropper links are comparatively weak) this statement may not hold, but this scenario is beyond the scope of this paper.
The maximum RSCP appears at about $\eta  = 15dB$ for the JRJS strategy using $\lambda  = {3 \mathord{\left/
 {\vphantom {3 4}} \right.
 \kern-\nulldelimiterspace} 4}$, while it is $\eta  = 10dB$ for the TBRS strategy. Furthermore, as expected, increasing the number of available relays and jamming nodes will always be able to improve the reliability and security performance. However, the continued boosting of the jammer's power (decreasing $\lambda $) will not always improves the overall performance, because the interference improves initially the security, but then it starts to reduce the reliability, as $\lambda $ decreases. This further motivates the designer to carefully take into account the power sharing between relaying and jamming. The effect of the rate pair setting on the security and reliability of the JRJS strategy is neglected here, which follows a similar trend to that of the TBRS strategy.

\begin{figure}
\begin{center}
  \includegraphics[width=3in,height=3in,angle=0]{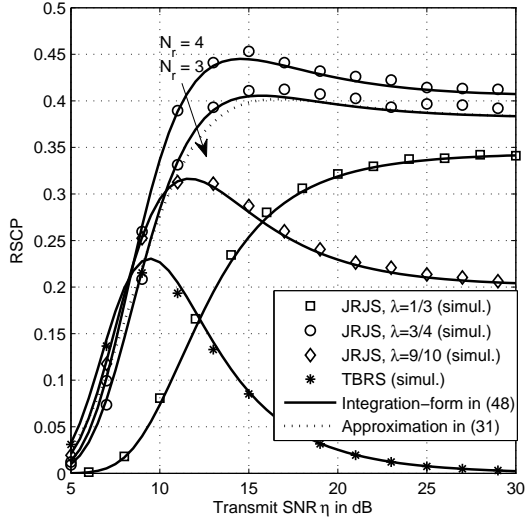}\\
  \caption{RSCP versus transmit SNR for the JRJS strategy for different power sharing ratios $\lambda$ as well as for $N_t  = K_r  = 3$, $f_d T_d  = 0.1$, and $R_0  = 1,R_s  = R_0 /8$.}
\end{center}
\end{figure}

Fig. 6 characterizes the RSR versus feedback delay and power sharing ratio for both TBRS and JRJS, in which the RSR curves are plotted by using (17) and (28), respectively. The first illustration shows that the RSR decreases as the delay coefficients ($\rho _{SR} $ and $\rho _{RD} $), which confirms that the improvement of reliability becomes more pronounced than the reduction of the security as the feedback delay decreases. This observation implies an improvement in terms of the security-reliability tradeoff. Besides, the RSR versus $\rho _{RD} $ is larger than that of $\rho _{SR} $, which indicates that the impact of the second-hop CSI feedback delay is more prominent. The other illustration in the right demonstrates that the RSR is a concave function of the power sharing ratio, which reflects the tradeoff between the reliability and security struck by adjusting $\lambda $.

\begin{figure}
\begin{center}
  \includegraphics[width=3in,height=3in,angle=0]{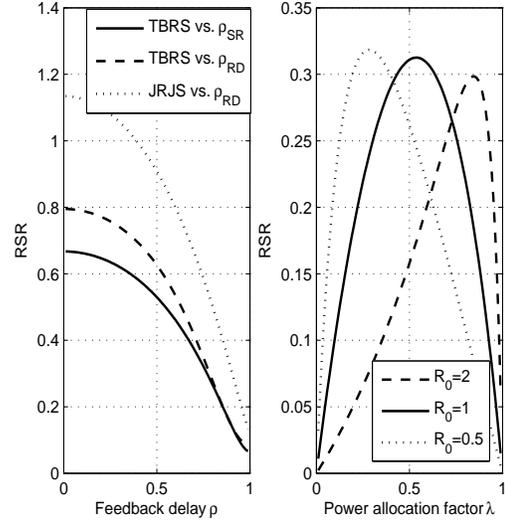}\\
  \caption{RSR versus feedback delay coefficient ($R_0  = 1,R_s  = R_0 /8, \lambda=3/4$) and power sharing ratio $\lambda $ ($R_s  = R_0 /8, \rho _{SR}  = \rho _{RD}  = 0.9$) for TBRS and JRJS strategy with $N_t  = K_r  = 3$.}
\end{center}
\end{figure}

\begin{figure}
\begin{center}
  \includegraphics[width=3in,height=3in,angle=0]{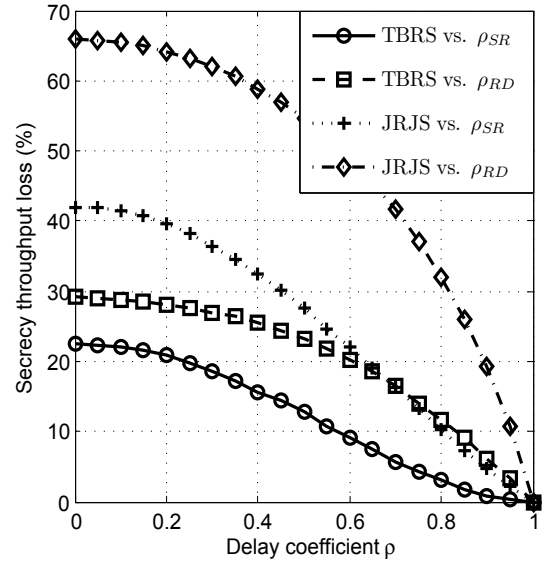}\\
  \caption{\emph{Percentage secrecy throughput loss versus delay coefficients with $N_t  = K_r  = 3$, $R_0  = 1,R_s  = R_0 /8$, $\lambda=3/4$ and $\eta  = 10dB$.}}
\end{center}
\end{figure}

To further evaluate the effect of feedback delays on the secrecy performance, Fig. 7 plots the resultant percentage of secrecy throughput loss versus the delay, which is defined as
\begin{equation}
\varsigma _{loss}  = \frac{{\varsigma _{no - delay}  - \varsigma _{delay} }}{{\varsigma _{no - delay} }}.
\end{equation}
It can be seen from the figure that compared to the TBRS scheme, JRJS is more sensitive to the feedback delays. Furthermore, recalling that increasing the delay coefficient $\rho_{SR}$ of the first hop improves the reliability, but at the same time it also helps the eavesdropper, hence it is not surprising that the secrecy throughput loss due to the second hop feedback delay is more pronounced.

\begin{figure}
\begin{center}
  \includegraphics[width=3in,height=3in,angle=0]{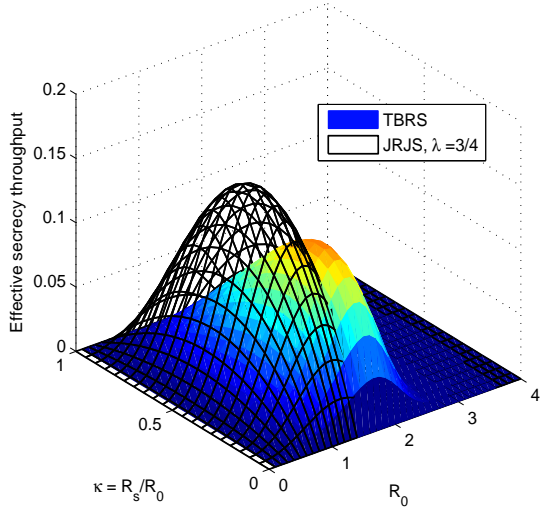}\\
  \caption{Secrecy throughput versus $R_0 $ and $ \kappa  = R_s/R_0 $ for both the TBRS and JRJS strategy with $N_t  = K_r  = 3$, $f_d T_d  = 0.1$ and $\eta  = 15{\rm{ d}}B$.}
\end{center}
\end{figure}

\begin{figure}
\begin{center}
  \includegraphics[width=3in,height=3in,angle=0]{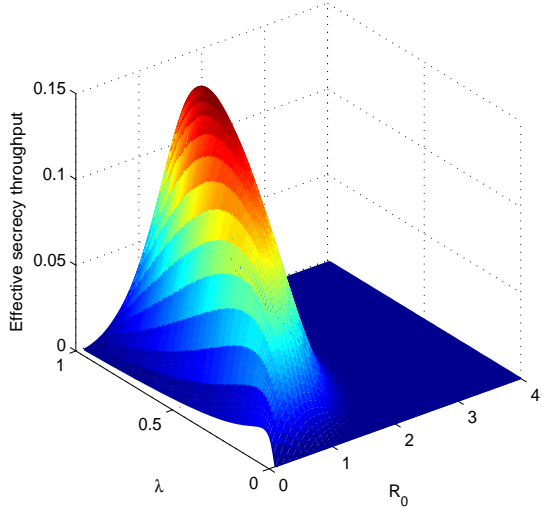}\\
  \caption{Secrecy throughput versus $R_0 $ and $\lambda $ for the JRJS strategy with $N_t  = K_r  = 3$, $f_d T_d  = 0.1$, $\eta  = 15{\rm{ d}}B$ and $R_s/R_0 = 1/8$.}
\end{center}
\end{figure}

Fig. 8 illustrates the achievable effective secrecy throughput for both the TBRS and JRJS strategies versus the codeword transmission rate $R_0 $ and the secrecy code ratio $\kappa  = {{R_s } \mathord{\left/
 {\vphantom {{R_s } {R_0 }}} \right.
 \kern-\nulldelimiterspace} {R_0 }}$ without any outage constraints ($\upsilon   = \delta  = 1$). The values of the effective secrecy throughput are plotted by using $\varsigma  = R_s P_{R\& S} $. We can observe in Fig. 8 that subject to a fixed code rate ratio, $\kappa $, the effective secrecy throughput increases to a peak value as $R_0 $ reaches its optimal value, and then decreases. This phenomenon can be explained as follows. At a low transmission rate, although the COP increases with $R_0 $, which has a negative effect on the effective secrecy throughput, both the secrecy rate and the SOP performance will benefit. However, after reaching the optimal $R_0 $, the effective secrecy throughput drops since the main link cannot afford a reliable transmission and the resultant COP increase becomes dominant. On the other hand, subject to a fixed $R_0 $ (which results in a constant COP), the effective secrecy throughput is also a concave function of $\kappa $, and increasing the code rate ratio ultimately results in an increased secrecy information rate at the cost of an increased SOP.

 The achievable effective secrecy throughput for the JRJS strategy is also presented in Fig. 8 and similar conclusions as well as trends can be observed to that of the TBRS case. Additionally, the comparison of the two strategies indicates that the JRJS scheme attains a higher effective secrecy throughput than the TBRS scheme operating without jamming, even if no power sharing optimization has been employed.

Fig. 9 further illustrates the impact of power sharing between the relay and the jammer on the achievable effective secrecy throughput of the JRJS strategy versus $R_0 $ in the absence of outage constraints. Given a fixed code rate pair $\left( {R_0 ,R_s } \right)$, the effective secrecy throughput follows the trend of the RSCP, which is a concave function of $\lambda $, as seen in Fig. 6. The interference introduced by the jammer initially improves both the reliability and security as $\lambda $ increases, but this trend is reversed beyond a certain point.

\section{Discussions}
\subsection{Impact of the S-E Link}
We note that the introduction of the S-E link, i.e. the information leakage in the first phase, is very critical to the security. There are also some researches focusing on the corresponding secure transmission design and performance evaluation for cooperative networks with the S-E link, such as [15-16]. In this subsection, we
assume that the eavesdropper can receive information directly from the
source in the first phase. Thus, following the steps in the prior
sections, for the TBRS and JRJS schemes, it is clear that the
SNR experienced at the eavesdropper should be rewritten as
\begin{equation}\label{36}
\tilde \gamma _E^\tau   = \gamma _{SE}  + \gamma _E^\tau,
\end{equation}
where $\gamma _{SE} = {{P_s \left| {{\bf{w}}_{opt} \left( {t\left|
        {T_{d_{SR} } } \right.} \right){\bf{h}}_{SE} \left( t \right)}
    \right|^2 } \mathord{\left/ {\vphantom {{P_s \left|
          {{\bf{w}}_{opt} \left( {t\left| {T_{d_{SR} } } \right.}
            \right){\bf{h}}_{SE} \left( t \right)} \right|^2 } {N_0
    }}} \right.  \kern-\nulldelimiterspace} {N_0 }}$ follows the
exponential distribution with the average value $\bar \gamma _{SE} $,
$\tau = \left\{ {TBRS,JRJS} \right\}$, and $\gamma _E^\tau $ has been
defined in (4) and (7).

Then, the corresponding SOP, RSCP and effective secrecy throughput
have to be reconsidered. Unfortunately, to the best of our knowledge,
it is a mathematically intractable problem to obtain closed-form
results for the related performance evaluations. Therefore, we
resorted to numerical simulations for further investigating the impact
of the S-E link. Fig. 10 compares the effective secrecy throughput of the
TBRS and JRJS shcemes both with and without considering the direct S-E
link. It becomes clear that the information leakage in the first phase
will lead to a severe security performance degradation, especially for
the JRJS scheme, which will no longer be capable of maintaining a
steady throughput at high SNRs. The reason for this trend is that
increasing the transmit SNR will help the eavesdropper in the presence
of the direct S-E link.

\begin{figure}
\begin{center}
  \includegraphics[width=3in,height=3in,angle=0]{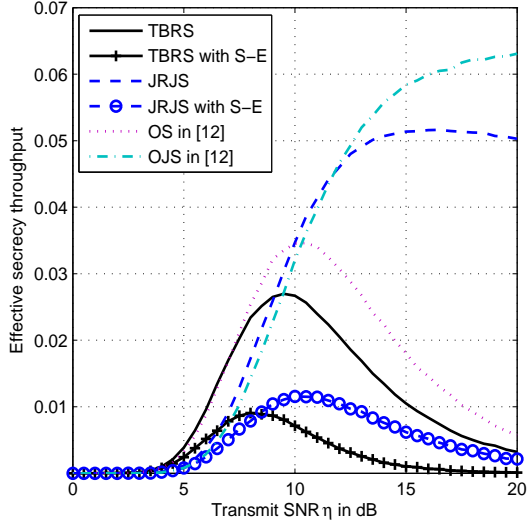}\\
  \caption{Comparisons for different strategies with and without the S-E link, for $N_t  = K_r  = 3$, $R_0  = 1,R_s  = R_0 /8$, $f_d T_d  = 0.1$, and $\lambda  = 3/4$.}
\end{center}
\end{figure}

\subsection{Comparisons}
In this subsection, based on the outdated CSI assumption, we provide
performance comparisons with a range of other schemes advocated in
[12] with the aid of the proposed outage-based
characterization. Fig. 10 also incorporates our effective secrecy
throughput performance comparison, where the optimal selection (OS)
regime as well as the optimal selection combined with jamming (OSJ)
were proposed in [12]. They are formulated as
\begin{equation}\label{37}
OS:R^*  = \arg \mathop {\max }\limits_{R_k  \in {\cal R}} \left\{ {{{\tilde \gamma _{R_k D} } \mathord{\left/
 {\vphantom {{\tilde \gamma _{R_k D} } {\tilde \gamma _{R_k E} }}} \right.
 \kern-\nulldelimiterspace} {\tilde \gamma _{R_k E} }}} \right\},
\end{equation}
and
\begin{equation}\label{38}
OSJ:\left\{ {\begin{array}{*{20}c}
   {R^*  = \arg \max _{R_k  \in {\cal R}} \left\{ {{{\tilde \gamma _{R_k D} } \mathord{\left/
 {\vphantom {{\tilde \gamma _{R_k D} } {\tilde \gamma _{R_k E} }}} \right.
 \kern-\nulldelimiterspace} {\tilde \gamma _{R_k E} }}} \right\}}  \\
   {J^*  = \arg \min _{R_k  \in {\cal R} - R^* } \left\{ {{{\tilde \gamma _{R_k D} } \mathord{\left/
 {\vphantom {{\tilde \gamma _{R_k D} } {\tilde \gamma _{R_k E} }}} \right.
 \kern-\nulldelimiterspace} {\tilde \gamma _{R_k E} }}} \right\}}  \\
\end{array}} \right.,
\end{equation}
where $\tilde \gamma _{R_k E} $ is the delayed version of the
instantaneous CSI of the R-E link.  It should be noted that this
constitutes an entirely new performance characterization of these
schemes from the perspective of the effective secrecy throughput. It
can be seen from Fig. 1 that the selection combined with jamming
outperforms the corresponding non-jamming techniques at high SNRs,
albeit this trend may no longer prevail at low SNRs. Besides, compared
to those selections relying on the average SNRs of the R-E link, the
optimal selections relying on the idealized simplifying assumptions of
having global CSI (OS and OSJ schemes) knowledge can only achieve
throughput gains at high SNRs due to the inevitable feedback delay.

\section{Conclusions}
An outage-based characterization of cooperative relay networks was provided in the face of CSI feedback delays. Two types of relaying strategies were considered, namely the traditionally best relay selection strategy as well as the joint relay and jammer selection strategy. Closed-form expressions of the connection outage probability, of the secrecy outage probability and of the reliable-and-secure connection probability as well as of the reliability-security-ratio were derived. The RSR results demonstrated that the reliability is improved more substantially than the security performance, when the CSI feedback delays is reduced. Furthermore, we presented a modified effective secrecy throughput definition and demonstrated that the JRJS strategy achieves a significant effective secrecy throughput gain over the TBRS strategy. The transmit SNR, the secrecy codeword rate setting as well as the power sharing ratio between the relay and jammer nodes play important roles in striking a balance between the reliability and security in terms of the secrecy throughput. The impact of the direct S-E link and the performance comparisons with other selection schemes were also included. Additionally, our results demonstrate that the JRJS is more sensitive to the feedback delays and that the secrecy throughput loss due to the second hop feedback delay is more pronounced than that of the first hop.

\section*{Appendix A}
\section*{Proof of Proposition 1}
In order to simplify the asymptotic performance analysis, (3) can be expressed in a more mathematically tractable form by the commonly used tight upper bound of $\gamma _D^{TBRS}  \le \min \left\{ {\gamma _{SR} ,\gamma _{R^* D} } \right\}$ and $\gamma _E^{TBRS}  \le \min \left\{ {\gamma _{SR} ,\gamma _{R^* E} } \right\}$. When we have $\eta  \to \infty $, based on the CDFs in (9) as well as (10) and closing the smallest order terms of ${x \mathord{\left/
 {\vphantom {x \eta }} \right.
 \kern-\nulldelimiterspace} \eta }$, we have
\begin{equation}\label{39}
\begin{array}{l}
 F_{\gamma _{SR} } \left( x \right) \!\!\to\!\! 1 \!\!-\!\! \!\left[\! {\sum\limits_{n \!=\! 0}^{N_t  \!-\! 1} {\!\!\!\left(\!\!\! {\begin{array}{*{20}c}
   {N_t  \!\!-\!\! 1}  \\
   n  \\
\end{array}} \!\!\right)\!\!\rho _{SR}^{2(N_t  \!-\! 1 \!-\! n)} (1 \!\!-\!\! \rho _{SR}^2 )^n } } \right. \!\!\!+\!\!\! \sum\limits_{n \!=\! 0}^{N_t  \!-\! 2} {\!\!\!\left(\!\!\! {\begin{array}{*{20}c}
   {N_t  \!\!-\!\! 1}  \\
   n  \\
\end{array}} \!\!\right)\!\!}  \\
 \left. { \times \rho _{SR}^{2(N_t  \!-\! 1 \!-\! n)} (1 \!-\! \rho _{SR}^2 )^n \frac{x}{{\bar \gamma _{SR} }} \!+\! {\cal O}\!\left(\! {\frac{x}{{\bar \gamma _{SR} }}} \!\right)\!} \right]\left[ {1 \!\!-\!\! \frac{x}{{\bar \gamma _{SR} }} \!+\! {\cal O}\!\left(\! {\frac{x}{{\bar \gamma _{SR} }}} \!\right)\!} \right] \\
  \!=\! 1 \!\!-\!\! \!\left[\! {1 \!+\! \!\left(\! {1 \!\!-\!\! \left( {1 \!-\! \rho _{SR}^2 } \right)^{N_t  \!-\! 1} } \right)\frac{x}{{\bar \gamma _{SR} }} \!\!+\!\! {\cal O}\!\left(\! {\frac{x}{{\bar \gamma _{SR} }}} \!\right)\!} \right]\!\left[\! {1 \!\!-\!\! \frac{x}{{\bar \gamma _{SR} }} \!+\! {\cal O}\!\left(\! {\frac{x}{{\bar \gamma _{SR} }}} \!\right)\!} \right] \\
  = \left( {1 - \rho _{SR}^2 } \right)^{N_t  - 1} \frac{x}{{\bar \gamma _{SR} }} + {\cal O}\left( {\frac{x}{{\bar \gamma _{SR} }}} \right) \\
 \end{array},
\end{equation}
where ${\cal O}\left( x \right)$ denotes the high-order infinitely small contributions as a function of $x$, and
\begin{equation}\label{40}
\begin{array}{l}
 F_{\gamma _{R^* D} } \left( x \right) \to 1 - \sum\limits_{k = 0}^{K_r  - 1} {( - 1)^k \frac{{K_r }}{{k + 1}}\left( {\begin{array}{*{20}c}
   {K_r  - 1}  \\
   k  \\
\end{array}} \right)}  \\
 ~~~~~~~~~~~~~~~ \times \!\left[\! {1 - \frac{{k + 1}}{{k\left( {1 - \rho _{RD}^2 } \right) + 1}}\frac{x}{{\bar \gamma _{RD} }} + {\cal O}\left( {\frac{x}{{\bar \gamma _{RD} }}} \right)} \!\right]\! \\
  = \sum\limits_{k = 0}^{K_r  \!-\! 1} {( - 1)^k \!\!\left(\!\! {\begin{array}{*{20}c}
   {K_r  \!\!-\!\! 1}  \\
   k  \\
\end{array}} \!\!\right)\!\!} \frac{{K_r }}{{k\left( {1 - \rho _{RD}^2 } \right) + 1}}\frac{x}{{\bar \gamma _{RD} }} \!+\! {\cal O}\left( {\frac{x}{{\bar \gamma _{RD} }}} \right) \\
 \end{array}.
\end{equation}

Then, applying the upper bound of the receiver SNR, we may rewrite the COP and SOP of the TBRS strategy at high SNRs as
\begin{equation}\label{41}
\begin{array}{l}
 P_{co}^{TBRS,\infty }  = 1 - \left( {1 - F_{\gamma _{SR^* } } \left( {\gamma _{th}^D } \right)} \right)\left( {1 - F_{\gamma _{R^* D} } \left( {\gamma _{th}^D } \right)} \right) \\
  \!\!=\!\! \!\left[\! {\frac{{\left( {1 \!-\! \rho _{SR}^2 } \right)^{N_t  \!-\! 1} }}{{\sigma _{SR}^2 }} \!\!+\!\! \sum\limits_{k \!=\! 0}^{K_r  \!-\! 1} {( \!-\! 1)^k \!\!\!\left(\!\!\! {\begin{array}{*{20}c}
   {K_r  \!\!-\!\! 1}  \\
   k  \\
\end{array}} \!\!\right)\!\!} \frac{{K_r }}{{\left[ {k\left( {1 \!-\! \rho _{RD}^2 } \right) \!+\! 1} \right]\sigma _{RD}^2 }}} \!\right]\!\frac{{2^{2R_0 }  \!-\! 1}}{\eta } \\
 \end{array}
\end{equation}
and according to the fact that $\gamma _{R^* E} $ is exponentially distributed, we have
\begin{equation}\label{42}
\begin{array}{l}
 1 \!\!-\!\! P_{so}^{TBRS,\infty }  \!\!=\!\! 1 \!\!-\!\! \left( {1 \!\!-\!\! F_{\gamma _{SR^* } } \left( {\gamma _{th}^E } \right)} \right)\left( {1 \!\!-\!\! F_{\gamma _{R^* E} } \left( {\gamma _{th}^E } \right)} \right) \\
  = \left[ {\frac{{\left( {1 - \rho _{SR}^2 } \right)^{N_t  - 1} }}{{\sigma _{SR}^2 }} + \frac{1}{{\sigma _{RE}^2 }}} \right]\frac{{2^{2\left( {R_0  - R_s } \right)}  - 1}}{\eta } \\
 \end{array}.
\end{equation}

Finally, substituting (41) and (42) into the definition of RST in (16), we can obtain (17).
\section*{Appendix B}
\section*{Proof of Lemma 1}
According to the description of COP and SOP, replacing $F_{\gamma _{R^* D} } \left( x \right)$ and $F_{\gamma _{R^* E} } \left( x \right)$ by $F_{\xi _D } \left( x \right)$ and $F_{\xi _E } \left( x \right)$ in (12) and (14) will involve a mathematically intractable integration of the form:
\begin{equation}\label{43}
\Upsilon \left( {a,b,\mu ,\nu } \right) = \int_0^\infty  {\frac{{z^a }}{{z + b}}\exp \left( { - \mu z - \frac{\nu }{z}} \right)dz},
\end{equation}
which, to the best of our knowledge, does not have a closed-form solution. Alternatively, bearing in mind that the integration above has a great matter with $\xi _D$, we now focus our attention on the approximation of $\xi _D$. Based on the PDF results in (23), it may be seen that $\gamma _{J^* D} $ obeys an exponential distribution. Then we can approximate $\hat \gamma _{J^* D}  = \gamma _{J^* D}  + 1$ also by the exponential distribution with an average value of $\mathbb{E}\left\{ {\hat \gamma _{J^* D} } \right\} = \frac{{\left[ {\left( {K_r  - 1} \right)(1 - \rho _{RD}^2 ) + 1} \right]\bar \gamma _{RD}  + K_r }}{{K_r }}$ by assuming that the AWGN term '1' is part of the stochastic mean terms. The approximation based on this method provides an very accurate analysis and the accuracy of this method is verified by the numerical results of [34]. Thus, the CDF of $\hat \xi _D  = {{\gamma _{R^* D} } \mathord{\left/
 {\vphantom {{\gamma _{R^* D} } {\hat \gamma _{J^* D} }}} \right.
 \kern-\nulldelimiterspace} {\hat \gamma _{J^* D} }}$ can be derived as
\begin{equation}\label{44}
F_{\hat \xi _D } \left( x \right) = \sum\limits_{k = 0}^{K_r  - 1} {( - 1)^k \!\left(\! {\begin{array}{*{20}c}
   {K_r  \!-\! 1}  \\
   k  \\
\end{array}} \!\right)\!\frac{{K_r }}{{k + 1}}} \frac{x}{{x + \hat \varphi _k }},
\end{equation}
where $\hat \varphi _k  = {{\mathbb{E}\left\{ {\gamma _{R^* D} } \right\}} \mathord{\mathbb{E}\left/
 {\vphantom {{\left\{ {\gamma _{R^* D} } \right\}} {\left\{ {\hat \gamma _{J^* D} } \right\}}}} \right.
 \kern-\nulldelimiterspace} {\left\{ {\hat \gamma _{J^* D} } \right\}}}$.

Then, substituting (44) into (11), we have
\begin{equation}\label{45}
\begin{array}{l}
 F_{\gamma _D^{JRJS} } \left( x \right) \!\!\approx\!\! \sum\limits_{n \!=\! 0}^{N_t  \!-\! 1} {\sum\limits_{k \!=\! 0}^{K_r  \!-\! 1} {\sum\limits_{m = 0}^{N_t  \!-\! 1 \!-\! n} {\!\!\left(\!\! {\begin{array}{*{20}c}
   {N_t  \!\!-\!\! 1}  \\
   n  \\
\end{array}} \!\right)\!\!\!\left(\!\! {\begin{array}{*{20}c}
   {K_r  \!\!-\!\! 1}  \\
   k  \\
\end{array}} \!\right)\!\!\!\left(\!\! {\begin{array}{*{20}c}
   {N_t  \!\!-\!\! 1 \!\!-\!\! n}  \\
   m  \\
\end{array}} \!\!\!\right)\!\!\!} } }  \\
 ~~~~~~~~~~~\times \frac{{( - 1)^k K_r \rho _{SR}^{2(N_t  \!-\! 1 \!-\! n)} \left( {1 \!-\! \rho _{SR}^2 } \right)^n \varphi _k x^{N_t  \!-\! 1 \!-\! n \!-\! m} e^{ - \frac{x}{{\bar \gamma _{SR} }}} }}{{\left( {N_t  - 1 - n} \right)!\left( {k + 1} \right)\bar \gamma _{SR}^{N_t  - n} \left( {x + \varphi _k } \right)}}\\
  ~~~~~~~~~~~\times \int_0^\infty  {\frac{{z^{m + 1} }}{{z + \frac{{x(x + 1)}}{{x + \varphi _k }}}}\exp \left( { - \frac{z}{{\bar \gamma _{SR} }}} \right)dz}  \\
 \end{array}.
\end{equation}

Using [33, Eq. (3.383.10)], we can obtain the CDF of $\gamma _D^{JRJS} $ as
\begin{equation}\label{46}
\begin{array}{l}
F_{\gamma _D^{JRJS} } \left( x \right) \!\! \approx\!\! 1 \!\!-\!\! \sum\limits_{n = 0}^{N_t  \!-\! 1} {\sum\limits_{k \!=\! 0}^{K_r\!-\!1 } {\sum\limits_{m = 0}^{N_t  \!-\! 1 \!-\! n} { \!\!\!\left(\!\!\! {\begin{array}{*{20}c}
   {N_t  \!\!-\!\! 1}  \\
   n  \\
\end{array}} \!\!\right)\!\!\!\!\left(\!\! {\begin{array}{*{20}c}
   {K_r\!\!-\!\!1 }  \\
   k  \\
\end{array}} \!\!\right)\!\!\!\!\left( \!\!{\begin{array}{*{20}c}
   {N_t  \!\!-\!\! 1 \!\!-\!\! n}  \\
   m  \\
\end{array}} \!\!\!\!\right)\!\!\!\!} } } \\
 \times \frac{( \!-\! 1)^k {\left( {K_r  + 1} \right)\rho _{SR}^{2(N_t  \!-\! 1 \!-\! n)} \left( {1 \!-\! \rho _{SR}^2 } \right)^n }}{{\left( {N_t  - 1 - n} \right)!\left( {k + 1} \right)\bar \gamma _{SR}^{N_t  - n} }}\frac{{\Gamma \left( {m \!+\! 2} \right)\hat \varphi _k x^{N_t  - n} \left( {x  + 1} \right)^{m + 1} }}{{\left( {x  \!+\! \hat \varphi _k } \right)^{m \!+\! 2} }} \\
 \times \exp \left[ { - \frac{{x \left( {\hat \varphi _k  - 1} \right)}}{{\bar \gamma _{SR} \left( {x  + \hat \varphi _k } \right)}}} \right]\Gamma \left( { - m - 1,\frac{{x (x  + 1)}}{{\bar \gamma _{SR} \left( {x  + \hat \varphi _k } \right)}}} \right) \\
 \end{array}.
\end{equation}

Finally, substituting $x = \gamma _{th}^D$ into (46), we obtain $P_{co}^{JRJS}$.

As far as the SOP is considered, we exploit the commonly used tight upper bound of $\gamma _E^{JRJS}  \ge \frac{1}{2}\min \left\{ {\gamma _{SR} ,\xi _E } \right\}$ to calculate it, which may be rewritten as
\begin{equation}\label{47}
\begin{array}{l}
P_{so}^{JRJS}  \approx \Pr \left\{ {\frac{1}{2}\min \left\{ {\gamma _{SR} ,\xi _E } \right\} > \gamma _{th}^E } \right\} \\
 ~~~~~~~~= \left[ {1 - F_{\gamma _{SR} } \left( {2\gamma _{th}^E } \right)} \right]\left[ {1 - F_{\xi _E } \left( {2\gamma _{th}^E } \right)} \right]\\
 \end{array}.
\end{equation}

Substituting (9) and (25) into (47), we obtain $P_{so}^{JRJS}$.

\section*{Appendix C}
\section*{Proof of Lemma 3}
According to the definition of the RSCP in (18), we can calculate it by
\begin{equation}\label{48}
\begin{array}{l}
 P_{R\& S}^{JRJS}  = \int_0^\infty  {\left[ {1 - F_{\xi _D } \left( {\gamma _{th}^D  + \frac{{\gamma _{th}^D \left( {\gamma _{th}^D  + 1} \right)}}{z}} \right)} \right]}  \\
  ~~~~~~~~~~~~~\times F_{\xi _E } \left( {\gamma _{th}^E  + \frac{{\gamma _{th}^E \left( {\gamma _{th}^E  + 1} \right)}}{{z + \gamma _{th}^D  - \gamma _{th}^E }}} \right)f_{\gamma _{SR^* } } \left( {z + \gamma _{th}^D } \right)dz \\
 \end{array}.
\end{equation}

In order to make the integration mathematically tractable, we invoke a simple approximation for $F_{\xi _E } \left( x \right)$
  by treating the AWGN term '1' in $\xi _E  = {{\gamma _{R^* E} } \mathord{\left/
 {\vphantom {{\gamma _{R^* E} } {\left( {\gamma _{J^* E}  + 1} \right)}}} \right.
 \kern-\nulldelimiterspace} {\left( {\gamma _{J^* E}  + 1} \right)}}$
 as part of the stochastic mean terms. Hence we have
\begin{equation}\label{49}
F_{\xi _E } \left( x \right) = \frac{x}{{x + \hat \phi }},
\end{equation}
where $\hat \phi  = \frac{{\lambda \eta \sigma _{RE}^2 }}{{\left( {1 - \lambda } \right)\eta \sigma _{RE}^2  + 1}}$.

Then, replacing the corresponding CDFs of the second hop with $F_{\hat \xi _D } \left( x \right)$ and $F_{\hat \xi _E } \left( x \right)$ in (26), the integration can be derived as
\begin{equation}\label{50}
\begin{array}{l}
 P_{R\& S}^{JRJS} \!\! \approx\!\! 1 \!\!\!-\!\!\! \sum\limits_{n \!=\! 0}^{N_t  \!-\! 1} {\sum\limits_{k \!=\! 0}^{K_r\!-\!1 } {\sum\limits_{m = 0}^{N_t  \!-\! 1 \!-\! n} { ( \!-\! 1)^k \!\!\!\left(\!\!\! {\begin{array}{*{20}c}
   {N_t  \!\!-\!\! 1}  \\
   n  \\
\end{array}} \!\!\right)\!\!\!\!\left(\!\! {\begin{array}{*{20}c}
   {K_r\!\!-\!\!1 }  \\
   k  \\
\end{array}} \!\!\right)\!\!\!\!\left( \!\!{\begin{array}{*{20}c}
   {N_t  \!\!-\!\! 1 \!\!-\!\! n}  \\
   m  \\
\end{array}} \!\!\!\!\right)\!\!\!\!} } } \\
 \frac{{\left( {K_r  \!+\! 1} \right)\rho _{SR}^{2(N_t  \!-\! 1 \!-\! n)} \!\left(\! {1 \!-\! \rho _{SR}^2 } \!\right)\!^n }}{{\left( {N_t  \!-\! 1 \!-\! n} \right)!\left( {k + 1} \right)\bar \gamma _{SR}^{N_t  - n} }}\frac{{\hat \varphi _k \left( {\gamma _{th}^D } \right)^{N_t  \!-\! 1 \!-\! n \!-\! m} }}{{\gamma _{th}^D  + \hat \varphi _k }}\exp \!\left(\! { \!-\! \frac{{\gamma _{th}^D }}{{\bar \gamma _{SR} }} \!-\! \frac{{\gamma _{th}^D }}{{\omega _k \bar \gamma _{RD} }}} \!\right)\! \\
\int_0^\infty  {e^{ \frac{-z}{{\bar \gamma _{SR} }}} z^{m \!+\! 1} \!\!\left[\!\! {\frac{1}{{z + \theta _{1,k} }} \!\!-\!\! \frac{{\hat \phi \left( {z+ \gamma _{th}^D  - \gamma _{th}^E } \right)e^{ \frac{{-\gamma _{th}^E }}{{\bar \gamma _{RE} }}} }}{{\left( {\gamma _{th}^E  + \hat \phi } \right)\left( {\theta _{1,k}  - \theta _2 } \right)}}\!\!\left(\!\! {\frac{1}{{z \!+\! \theta _2 }} \!\!-\!\! \frac{1}{{z \!+\! \theta _{1,k} }}} \!\!\right)\!\!} \!\right]\!} dz \\
 \end{array},
\end{equation}
where $\hat \varphi _k$ and $\hat \phi$ are introduced by relying on the similar approximation as in Appendix B. Then, using [33, Eq. (3.383.10)], we obtain $P_{R\& S}^{JRJS}$.

\ifCLASSOPTIONcaptionsoff
  \newpage
\fi




\begin{thebibliography}{1}

\bibitem{}  B. Schneier,``Cryptographic design vulnerabilities," \emph{Computer}, vol. 31, no. 9, pp. 29-33, Sep. 1998.
\bibitem{}  A. D. Wyner, ``The wire-tap channel," \emph{Bell Syst. Tech. J.}, vol. 54, no. 8, pp. 1355-1387, 1975.
\bibitem{}  I. Csiszar and J. Korner, ``Broadcast channels with confidential messages," \emph{IEEE Trans. Inform. Theory}, vol. 24, no. 3, pp. 339-348, May 1978.
\bibitem{}  W. K. Harrison, J. Almeida, M. R. Bloch, S. W. McLaughlin, and J. Barros, ``Coding for secrecy: An overview of error-control coding techniques for physical-layer security," \emph{IEEE Signal Processing Magazine}, vol. 30, no. 5, pp. 41-50, Sep. 2013.
\bibitem{}  P. K. Gopala, L. Lai, and H. E. Gamal, ``On the secrecy capacity of fading channels," \emph{IEEE Trans. Inf. Theory}, vol. 54, no. 10, pp. 4687- 4698, Oct. 2008.
\bibitem{}  Y. W. P. Hong, P. C. Lan, and C. C. J. Kuo, ``Enhancing physical-layer secrecy in multi-antenna wireless systems: An overview of signal processing approaches," \emph{IEEE Signal Processing Magazine}, vol. 30, no. 5, pp. 29-40, Sep. 2013.
\bibitem{}  R. Bassily, E. Ekrem, X. He, E. Tekin, J. Xie, M. R. Bloch, S. Ulukus, and A. Yener, ``Cooperative security at the physical layer: A summary of recent advances," \emph{IEEE Signal Processing Magazine}, vol. 30, no. 5, pp. 16-28, Sep. 2013.
\bibitem{}  L. Dong, Z. Han, A. P. Petropulu, and H. V. Poor, ``Improving wireless physical layer security via cooperating relays," \emph{IEEE Trans. Signal Process.}, vol. 58, no. 3, pp. 1875-1888, Mar. 2010.
\bibitem{}  J. Huang and A. L. Swindlehurst, ``Cooperative jamming for secure communications in MIMO relay networks," \emph{IEEE Trans. Signal Process.}, vol. 59, no. 10, pp. 4871-4884, Oct. 2011.
\bibitem{}  Y. Zou, X. Wang, and W. Shen, ``Optimal relay selection for physical-layer security in cooperative wireless networks," \emph{IEEE J. Sel. Areas Commun.}, vol. 31, no. 10, pp. 2099-2111, Oct. 2013.
\bibitem{}  Y. Zou, X. Wang, W. Shen, and L. Hanzo, ``Security versus reliability analysis of opportunistic relaying," \emph{IEEE Trans. Veh. Tech.}, vol. 63, no. 6, pp. 2653-2661, June 2014.
\bibitem{}  I. Krikidis, J. S. Thompson, and S. McLaughlin, ``Relay selection for secure cooperative networks with jamming," \emph{IEEE Trans. Wireless Commun.}, vol. 8, no. 10, pp. 5003-5011, Oct. 2009.
\bibitem{}  J. Chen, R. Zhang, L. Song, Z. Han, and B. Jiao, ``Joint relay and jammer selection for secure two-way relay networks," \emph{IEEE Trans. Inf. Foren. Sec.}, vol. 7, no. 1, pp. 310-320, Feb. 2012.
\bibitem{}  Z. Ding, M. Xu, J. Lu, and F. Liu, ``Improving wireless security for bidirectional communication scenarios," \emph{IEEE Trans. Veh. Tech.}, vol. 61, no. 6, pp. 2842-2848, Oct. 2012.
\bibitem{}  C. Wang, H. M. Wang, and X. G. Xia, ``Hybrid opportunistic relaying and jamming with power allocation for secure cooperative networks," \emph{IEEE Trans. Wireless Commun.}, vol. 14, no. 2, pp. 589-605, 2015.
\bibitem{}  H. Deng, H. M. Wang, W. Guo, and W. Wang, ``Secrecy transmission with a helper: to relay or to jam," \emph{IEEE Trans. Inf. Foren. Sec.}, vol. 10, no. 2, pp. 293-307, 2015.
\bibitem{}  B. He, X. Zhou, and T. D. Abhayapala, ``Wireless physical layer security with imperfect channel state information: A survey," \emph{ZTE Communications}, vol. 11, no. 3, pp. 11-19, Sep. 2013.
\bibitem{}  A. Mukherjee and A. L. Swindlehurst, ``Robust beamforming for security in MIMO wiretap channels with imperfect CSI," \emph{IEEE Trans. Signal Process.}, vol. 59, no. 1, pp. 351-361, Jan. 2011.
\bibitem{}  J. Zhang and M. C. Gursoy, ``Relay beamforming strategies for physical-layer security," \emph{ in Proc. }, CISS, Princeton, NJ, Mar. 2010, pp. 1-6.
\bibitem{}  M. Bloch, J. Barros, M. R. D. Rodrigues, and S. W. McLaughlin, ``Wireless information-theoretic security," \emph{IEEE Trans. Inf. Theory}, vol. 54, no. 6, pp. 2515-2534, June 2008.
\bibitem{}  X. Zhou, M. R. McKay, B. Maham, and A. Hjorungnes, ``Rethinking the secrecy outage formulation: A secure transmission design perspective," \emph{IEEE Commun. Lett.}, vol. 15, no. 3, pp. 302-304, Mar. 2011.
\bibitem{}  J. Hu, Y. Cai, N. Yang, and W. Yang, ``A new secure transmission scheme with outdated antenna selection," \emph{IEEE Trans. Inf. Forensics Security}, accepted to appear.
\bibitem{}  J. Hu, W. Yang, N. Yang, X. Zhou and Y. Cai, ``On-off-based secure transmission design with outdated channel state information," \emph{IEEE Trans. Veh. Technol.}, accepted to appear.
\bibitem{}  N. E. Wu and H. J. Li, ``Effect of feedback delay on secure cooperative networks with joint relay and jammer selection," \emph{IEEE Wireless Commun. Lett.}, vol. 2, no. 4, pp. 415-418, July 2013.
\bibitem{}  X. Guan Y. Cai and Y. Yang, ``Secure Transmission design and performance analysis for cooperation exploring outdated CSI," \emph{IEEE Commun. Lett.}, vol. 18, no. 9, pp. 1637-1640, 2014.
\bibitem{}  L. Wang, S. Xu, W. Yang, W. Yang, and Y. Cai, ``Security performance of multiple antennas multiple relaying networks with outdated relay selection," \emph{in Proc.}, WCSP 2014, Hefei, China, Oct 2014, pp. 1-6.
\bibitem{}  J. Huang and A. Lee Swindlehurst, ``Buffer-aided relaying for two-hop secure communication," \emph{IEEE Trans. Wireless Commun.}, vol. 14, no. 1, pp. 152-164, 2015.
\bibitem{}  S. I. Kim, I. M. Kim, and J. Heo, ``Secure transmission for multiuser relay networks," \emph{IEEE Trans. Wireless Commun.}, Accepted, 0.1109/TWC. 2015. 2410776.
\bibitem{}  Y. Ma, D. Zhang, A. Leith, and Z. Wang, ``Error performance of transmit beamforming with delayed and limited feedback," \emph{IEEE Trans. Wireless Commun.}, vol. 8, no. 3, pp. 1164-1170, Mar. 2009.
\bibitem{}  Z. Rezki, A. Khisti, and M. S. Alouini, ``Ergodic secret message capacity of the wirchannel with finite-rate feedback," \emph{IEEE Trans. Wireless Commun.}, vol. 13, no. 6, pp. 3364-3379, 2014.
\bibitem{} X. Tang, R. Liu, P. Spasojevic, and H. V. Poor, "On the throughput of secure hybrid-ARQ protocols for Gaussian block-fading channels," \emph{IEEE Trans. Inf. Theory}, vol. 55, no. 4, pp. 1575-1591, Apr. 2009.
\bibitem{}  H. A. Suraweera, M. Soysa, C. Tellambura, and H. K. Garg, ``Performance analysis of partial relay selection with feedback delay," \emph{IEEE Signal Process. Lett.}, vol. 17, no. 6, pp. 531-534, Jun. 2010.
\bibitem{}  I. S. Gradshteyn and I. M. Ryzhik, \emph{Table of Integrals, Series and Products} 6th ed., San Diego: CA, Academic Press, 2000.
\bibitem{}  S. Kim and J. Heo. ``Outage Probability of Interference-Limited Amplify-and-Forward Relaying with Partial Relay Selection,"  \emph{in Proc.}, IEEE VTC, Yokohama, Japan, May 2011, pp. 1-5.

\end{thebibliography}
%

\begin{IEEEbiography}[{\includegraphics[width=1in,height=1.25in,clip,keepaspectratio]
{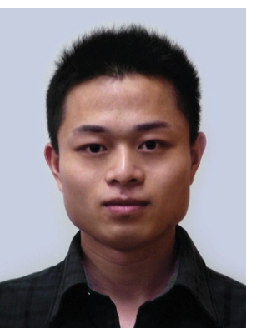}}]{Lei Wang} (S'11) received the B.S. degree in Electronics and Information Engineering from Central South University, Changsha, China in 2004, the M.S. degree in Communications and Information Systems from PLA University of Science and Technology, Nanjing, China in 2011. He is currently pursuing for the Ph.D degree in Communications and Information Systems at PLA University of Science and Technology. His current research interest includes cooperative communications, signal processing in communications, and physical layer security.
\end{IEEEbiography}

\begin{IEEEbiography}[{\includegraphics[width=1in,height=1.25in,clip,keepaspectratio]
{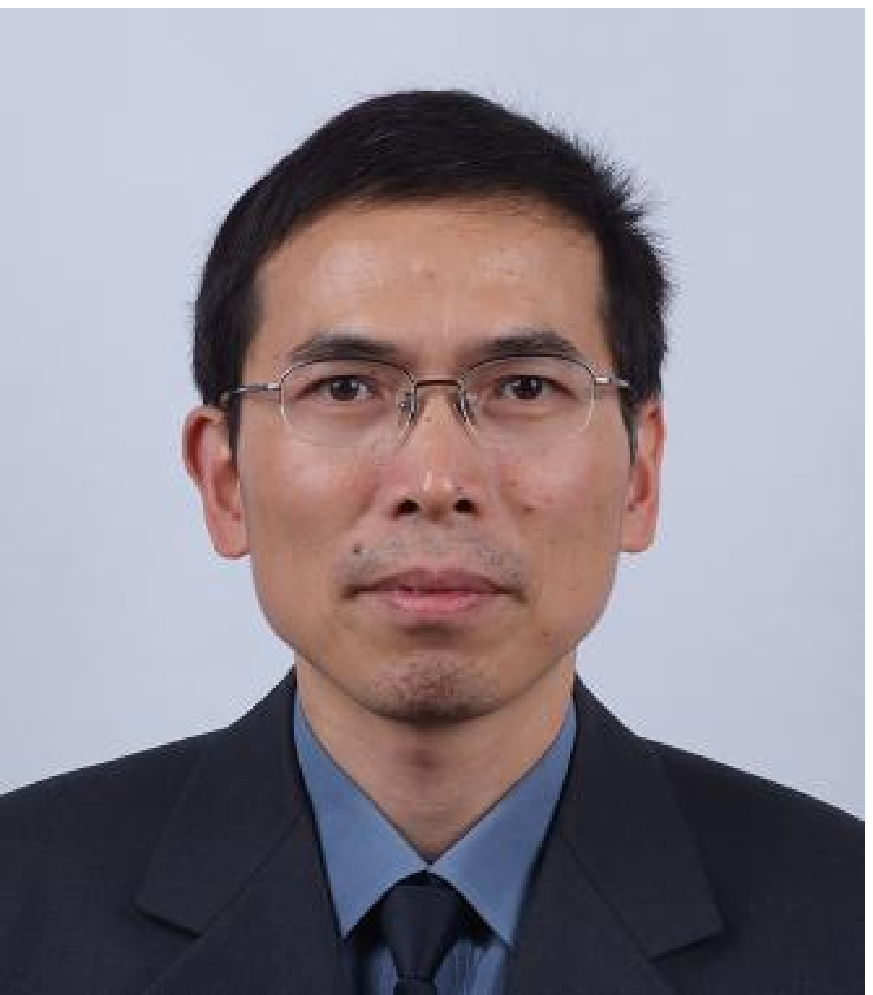}}]{Yueming Cai} (M'05-SM'12) received his B.S. degree in Physics from Xiamen University, Xiamen, China in 1982, the M.S. degree in Micro-electronics Engineering and the Ph.D. degree in Communications and Information Systems both from Southeast University, Nanjing, China in 1988 and 1996 respectively. His current research interest includes MIMO systems, OFDM systems, signal processing in communications, cooperative communications and wireless sensor networks.
\end{IEEEbiography}

\begin{IEEEbiography}[{\includegraphics[width=1in,height=1.25in]{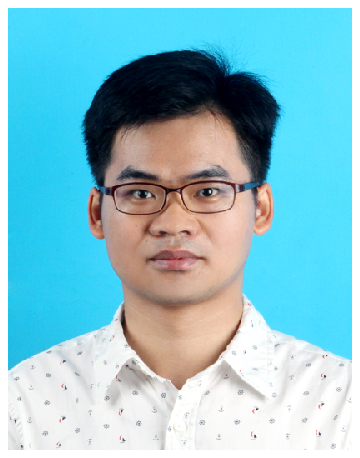}}]{Yulong Zou} (SM'13) is a Professor at the Nanjing University of Posts and Telecommunications (NUPT), Nanjing, China. He received the B.Eng. degree in Information Engineering from NUPT, Nanjing, China, in July 2006, the first Ph.D. degree in Electrical Engineering from the Stevens Institute of Technology, New Jersey, the United States, in May 2012, and the second Ph.D. degree in Signal and Information Processing from NUPT, Nanjing, China, in July 2012. His research interests span a wide range of topics in wireless communications and signal processing, including the cooperative communications, cognitive radio, wireless security, and energy-efficient communications.

Dr. Zou is a recipient of the 2014 IEEE Communications Society Asia-Pacific Best Young Researcher. He serves on the editorial board of the IEEE Communications Surveys and Tutorials, IEEE Communications Letters, IET Communications, and EURASIP Journal on Advances in Signal Processing. In addition, he has acted as symposium chairs, session chairs, and TPC members for a number of IEEE sponsored conferences, including the IEEE Wireless Communications and Networking Conference (WCNC), IEEE Global Communications Conference (GLOBECOM), IEEE International Conference on Communications (ICC), IEEE Vehicular Technology Conference (VTC), International Conference on Communications in China (ICCC), and so on.
\end{IEEEbiography}

\begin{IEEEbiography}[{\includegraphics[width=1in,height=1.25in,clip,keepaspectratio]
{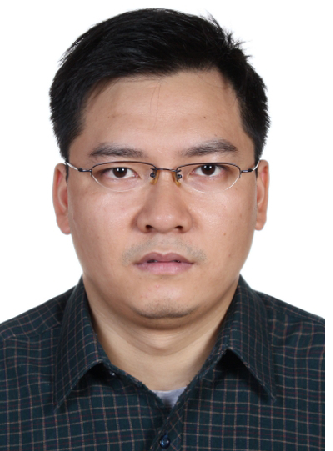}}]{Weiwei Yang} (S'08-M'12) received the B.S., M.S. and Ph.D. degrees from College of Communications Engineering, PLA University of Science and Technology, Nanjing, China, in 2003, 2006 and 2011, respectively. His research interests are orthogonal frequency domain multiplexing systems, signal processing in communications, cooperative communications, cognative networks and network security.
\end{IEEEbiography}

\begin{IEEEbiography}[{\includegraphics[width=1in,height=1.25in,clip,keepaspectratio]
{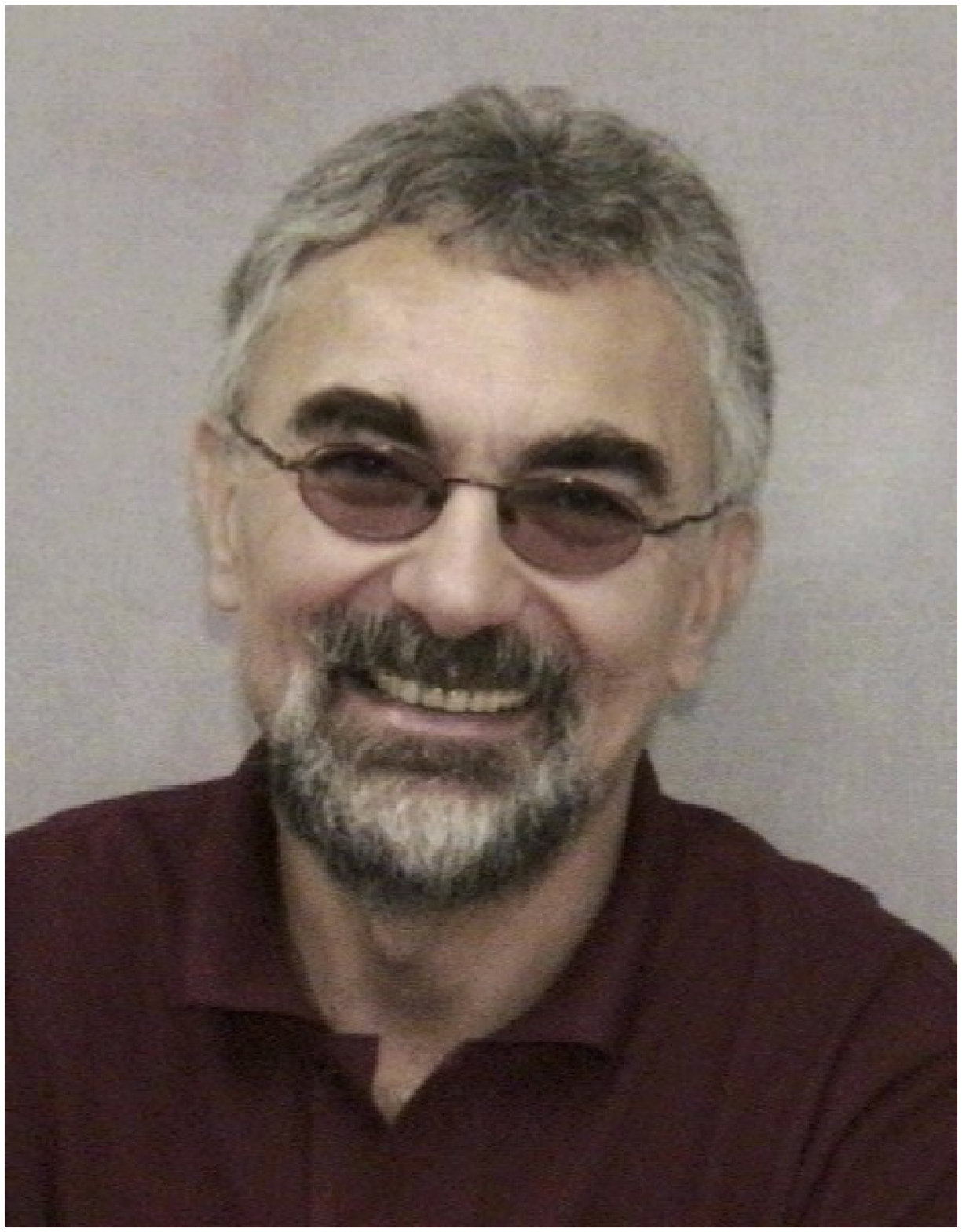}}]{\bf Lajos Hanzo} (http://www-mobile.ecs.soton.ac.uk) FREng, FIEEE, FIET, Fellow of EURASIP, DSc received his degree in electronics in 1976 and his doctorate in 1983.  In 2009 he was awarded an honorary doctorate by the Technical University of Budapest, while in 2015 by the University of Edinburgh.  During his 38-year career in telecommunications he has held various research and academic posts in Hungary, Germany and the UK. Since 1986 he has been with the School of Electronics and Computer Science, University of Southampton, UK, where he holds the chair in telecommunications.  He has successfully supervised about 100 PhD students, co-authored 20 John Wiley/IEEE Press books on mobile radio communications totalling in excess of 10 000 pages, published 1500+ research entries at IEEE Xplore, acted both as TPC and General Chair of IEEE conferences, presented keynote lectures and has been awarded a number of distinctions. Currently he is directing a 60-strong academic research team, working on a range of research projects in the field of wireless multimedia communications sponsored by industry, the Engineering and Physical Sciences Research Council (EPSRC) UK, the European Research Council's Advanced Fellow Grant and the Royal Society's Wolfson Research Merit Award. He is an enthusiastic supporter of industrial and academic liaison and he offers a range of industrial courses.  He is also a Governor of the IEEE VTS.  During 2008 - 2012 he was the Editor-in-Chief of the IEEE Press and a Chaired Professor also at Tsinghua University, Beijing.  His research is funded by the European Research Council's Senior Research Fellow Grant.  For further information on research in progress and associated publications please refer to http://www-mobile.ecs.soton.ac.uk Lajos has 22 000+ citations.
\end{IEEEbiography}

\end{document}